# Covalent Functionalized Black Phosphorus Quantum Dots

Francesco Scotognella[1,2,*], Ilka Kriegel[3,4], Simone Sassolini[4]
[1]*Dipartimento di Fisica, Politecnico di Milano, Piazza Leonardo da Vinci 32, 20133 Milano, Italy*
[2]*Center for Nano Science and Technology@PoliMi, Istituto Italiano di Tecnologia, Via Giovanni Pascoli, 70/3, 20133, Milan, Italy*
[3]*Department of Nanochemistry, Istituto Italiano di Tecnologia (IIT), via Morego, 30, 16163 Genova, Italy*
[4]*Molecular Foundry, Lawrence Berkeley National Lab, One Cyclotron Rd, Berkeley, CA 94720, USA*
*email address: francesco.scotognella@polimi.it

## Abstract

Black phosphorus (BP) nanostructures enable a new strategy to tune the electronic and optical properties of this atomically thin material. In this paper we show, via density functional theory calculations, the possibility to modify the optical properties of BP quantum dots via covalent functionalization. The quantum dot selected in this study has chemical formula P24H12 and has been covalent functionalized with one or more benzene rings or anthracene. The effect of functionalization is highlighted in the absorption spectra, where a red shift of the absorption is noticeable. The shift can be ascribed to an electron delocalization in the black phosphorus/organic molecule nanostructure.

## Introduction

The elemental two-dimensional analogues of graphene, i.e. germanene, phosphorene, silicene, and stanene, have emerged as a class of fascinating nanomaterials useful for optics and electronics [1–4]. Black phosphorus (BP) is particularly interesting because of its electronic band gap, which is 0.3 eV in the bulk, but that shows a layer-dependent tunability up to 2 eV [5–12]. From an optical point of view, a strong photoluminescence [8,13] and in-plane anisotropy [14] have been observed in BP. The integration of black phosphorus in one-dimensional photonic crystals and microcavities has been also presented [15].

An interesting treatment to BP is the covalent functionalization. Li et al. studied extensively with a first-principle approach the covalent functionalization of BP [16]. Ryder et al. experimentally demonstrated the covalent functionalization and passivation of exfoliated BP [17]. BP nanostructures enable a new strategy to tune the properties of the material. The first appearance of BP quantum dots is reported in 2015 [18,19]. The size dependence of absorption and emission of BP quantum dots has been studied [20] and show a peculiar behaviour, with the absorption gap following an inversely proportional law to the dot diameter, while the emission wavelength follows a mixed behaviour: proportional to the size up to 1.8 nm and inversely proportional above 1.8 nm. Moreover, the suitability of phosphorene quantum dots for photocatalysis has been discussed [21], while phosphorene quantum dot – fullerene composites have been suggested for solar energy conversion [22].

In this paper we study the possibility to tune the optical properties of BP quantum dots via covalent functionalization. The chosen quantum dot has chemical formula P24H12 and has been covalently functionalized with one or more benzene rings or anthracene. The effect of such functionalization with aromatic hydrocarbons is highlighted in the absorption spectra, where a red shift of the absorption is noticeable. The shift can be ascribed to electron delocalization in the nanostructure.

## Methods

The functionalized BP quantum dots have been designed with the Avogadro package [23]. The optimization of the ground state geometry and the calculation of the electronic transitions have been performed with the package ORCA 3.0.3 [24], using the B3LYP functional [25] in the framework of the density functional theory. The Ahlrichs split valence basis set [26] and the all-electron nonrelativistic basis set SVPalls1 [27,28] have been employed. Moreover, the calculation utilizes the Libint library [29].

## Results and Discussion

In Figure 1 we show the optimized geometry of the P24H12 BP quantum dot and three different functionalized quantum dots, with one, two, and four benzene rings. We report the coordinates of such geometries in the Supporting Information.

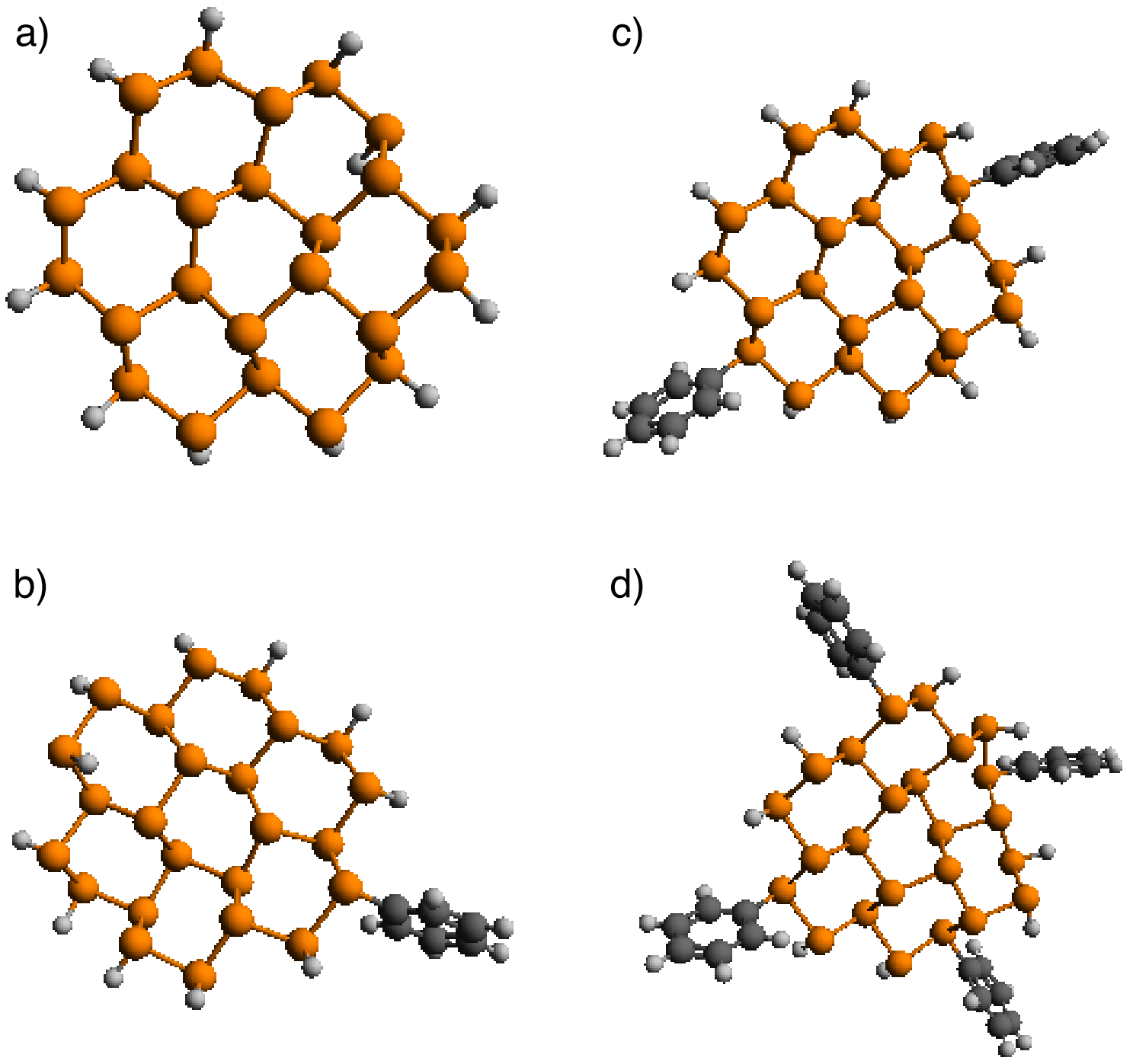

**Figure 1.** Optimized ground state geometry of (a) the P24H12 dot and of the benzene functionalized dots: (b) BenzP24H11; (c) 2BenzP24H10; (d) 4BenzP24H8.

We studied the electronic transitions of the four different BP quantum dots with density functional theory calculations and we report the absorption spectra in Figure 2. The transitions with the corresponding oscillator strengths are reported in the Supporting Information. In the simulated absorption spectra we have considered a spectral linewidth of 500 $cm^{-1}$. The bare quantum dots P24H12 are given in black (Figure 2) and show several resonances dominating the spectrum with an absorption onset around 420 nm. We remark here that we observe a slightly different energy position than in Ref. [20] which is most probably due to a different optimization of the geometry. The functionalization of P24H12 with a benzene ring (we refer to this structure as BenzP24H11, red curve in Figure 2) leads to a mild red shift of 1 nm of the first transition, while the higher energy transitions seem more influenced. Instead, the spectra of the BP quantum dot with two and four benzene rings (2BenzP24H10 and 4BenzP24H8, respectively) are significantly affected by the presence of the benzene rings, with a red shift of the first transition of 4BenzP24H8 by 7 nm (green curve in Figure 2) with a remarkable strong peak at 397 nm appearing for 4BenzP24H8. This can be ascribed to a wavefunction delocalization in the black phosphorus/organic molecule hybrid system. In Figure 3 we report the LUMO orbitals for P24H12 and 4BenzP24H8 and the spreading of the wavefunction over the benzene rings in 4BenzP24H8 is clearly visible. In the Supporting Information we reported the orbitals involved in the lowest transition.

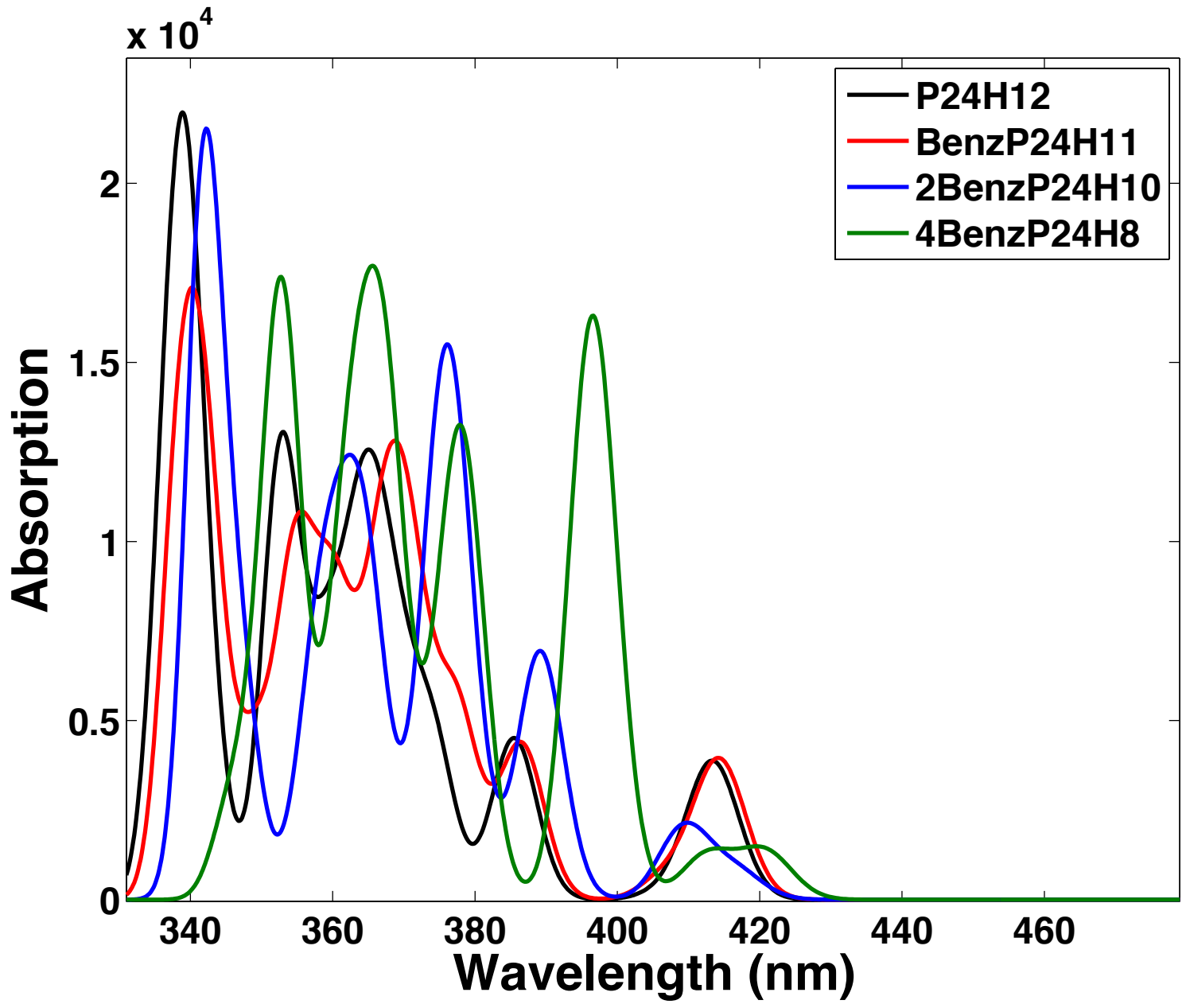

**Figure 2.** Absorption spectra of P24H12 (black curve), BenzP24H11 (red curve), 2BenzP24H10 (blue curve), and 4BenzP24H8 (green curve).

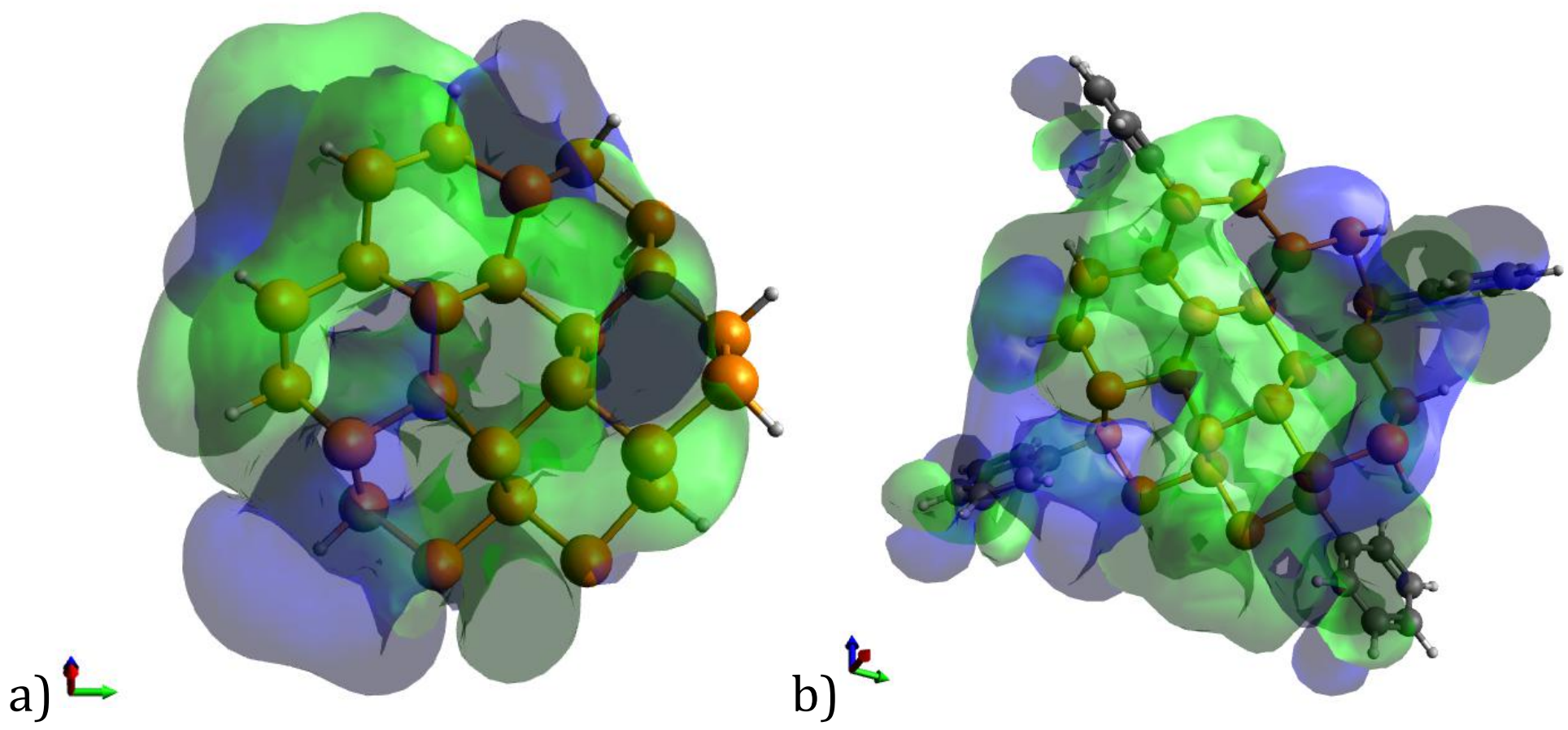

**Figure 3.** LUMO orbitals for (a) P24H12 and (b) 4BenzP24H8.

To study further the role of wavefunction delocalization we further studied the covalent functionalization of P24H12 with anthracene, an aromatic molecule composed of three benzene rings. The optimized geometry is shown in Figure 4 and highlights the planar structure of this molecule due to the aromatic nature. The coordinates are reported in the Supporting Information.

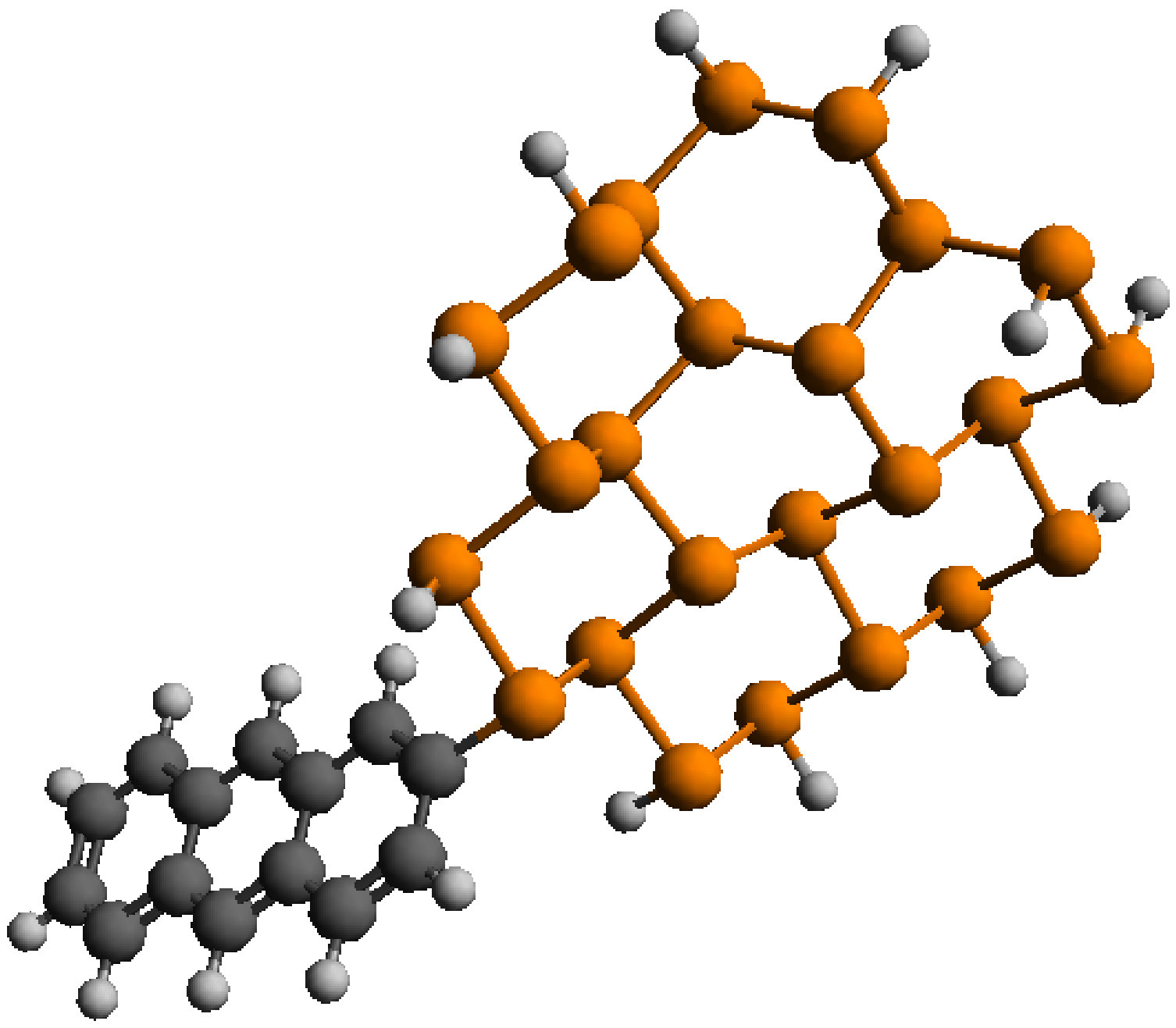

**Figure 4.** Anthracene functionalized BP quantum dot optimized geometry (AnthP24H11).

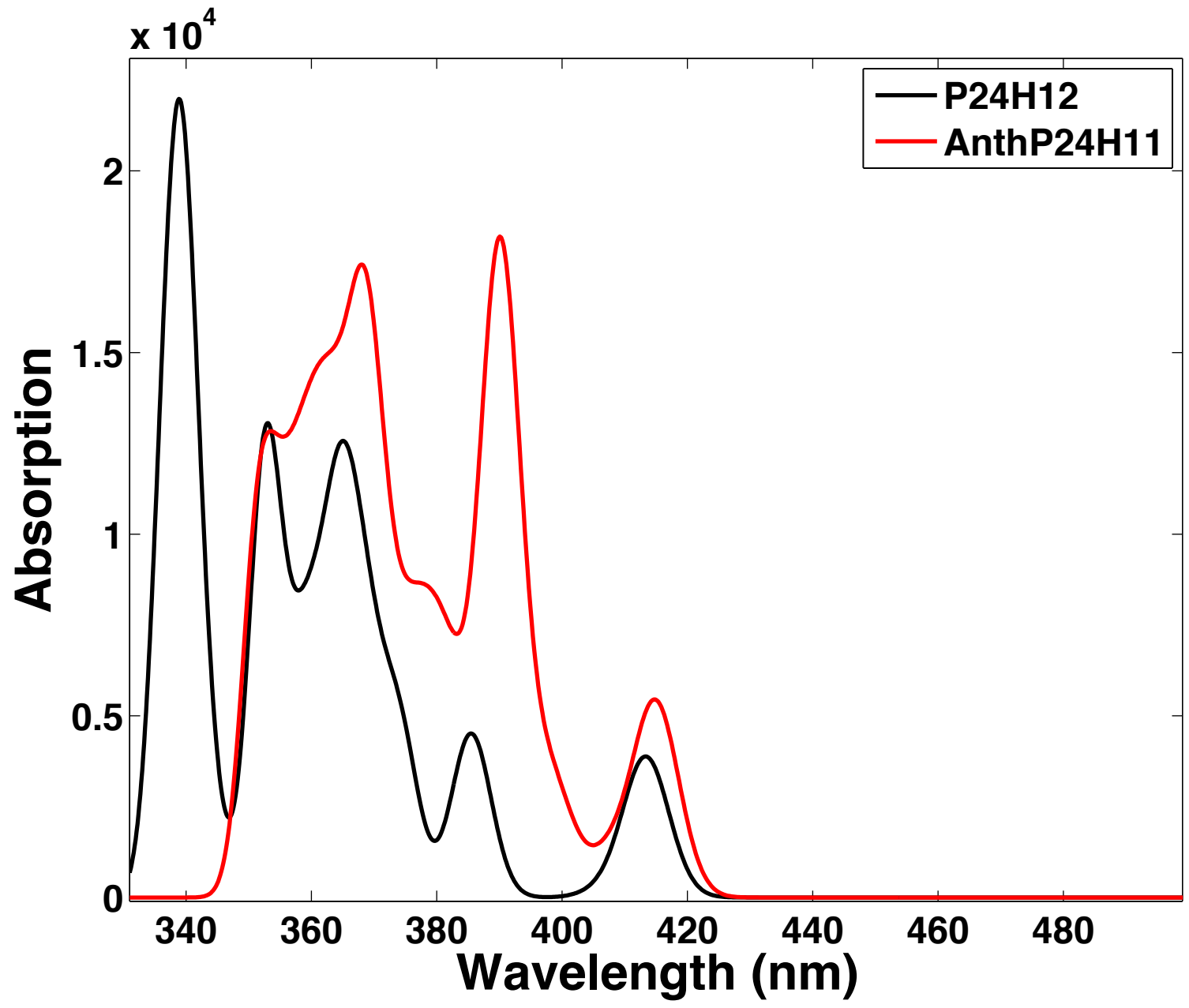

**Figure 5.** Absorption of the P24H12 BP quantum dot and the anthracene functionalized BP quantum dot (AnthP24H11).

We report in Figure 5 the absorption spectra of P24H12 (black curve) and the BP quantum dot functionalized with anthracene (AnthP24H11, red curve). A mild red shift of 1.6 nm is noticeable for the first transition. To underline that this shift is not to a superposition of the P24H12 absorption and the anthracene one we report the calculations of the anthracene transitions in the Supporting Information (the optimized ground state geometry of anthracene is taken from Ref. [30]).

The influence of benzene and anthracene covalent functionalization of BP quantum dots is clear and become more pronounced by increasing the number of molecules attached to the BP dot, with a stronger red shift of the absorption. Such phenomenon can be ascribed to an electron delocalization in the black phosphorus/organic molecule hybrid compound.

## Conclusion

In this work we have studied the opportunity to tailor the optical properties of BP quantum dots through a covalent functionalization with aromatic molecules. The selected BP quantum dot are made of 24 P atoms and 12 H atoms. It has been covalently functionalized with one or more benzene rings or anthracene. The effect of such functionalization is evident in the absorption spectra, where a red shift of the absorption is discerned. With a BP dot functionalization with four benzene rings, a red shift of 7 nm of the first transition has been observed. The red shift can be ascribed to a to an electron delocalization in the hybrid black phosphorus/organic molecule nanostructure.

## Acknowledgements

This project has received funding from the European Union's Horizon 2020 research and innovation programme (MOPTOPus) under the Marie Skłodowska-Curie grant agreement No. [705444], as well as (SONAR) grant agreement no. [734690]. Work at the Molecular Foundry was supported by the Office of Science, Office of Basic Energy Sciences, of the U.S. Department of Energy under Contract No. DE-AC02-05CH11231.

## References

[1] D. Tománek, Interfacing graphene and related 2D materials with the 3D world, J. Phys. Condens. Matter. 27 (2015) 133203. doi:10.1088/0953-8984/27/13/133203.

[2] M. Houssa, A. Dimoulas, A. Molle, Silicene: a review of recent experimental and theoretical investigations, J. Phys. Condens. Matter. 27 (2015) 253002. doi:10.1088/0953-8984/27/25/253002.

[3] A. Molle, J. Goldberger, M. Houssa, Y. Xu, S.-C. Zhang, D. Akinwande, Buckled two-dimensional Xene sheets, Nat. Mater. 16 (2017) 163–169. doi:10.1038/nmat4802.

[4] S. Balendhran, S. Walia, H. Nili, S. Sriram, M. Bhaskaran, Elemental Analogues of Graphene: Silicene, Germanene, Stanene, and Phosphorene, Small. 11 (2015) 640–652. doi:10.1002/smll.201402041.

[5] H.O.H. Churchill, P. Jarillo-Herrero, Two-dimensional crystals: Phosphorus joins the family, Nat. Nanotechnol. 9 (2014) 330–331. doi:10.1038/nnano.2014.85.

[6] L. Li, Y. Yu, G.J. Ye, Q. Ge, X. Ou, H. Wu, D. Feng, X.H. Chen, Y. Zhang, Black phosphorus field-effect transistors, Nat. Nanotechnol. 9 (2014) 372–377. doi:10.1038/nnano.2014.35.

[7] H. Liu, Y. Du, Y. Deng, P.D. Ye, Semiconducting black phosphorus: synthesis, transport properties and electronic applications, Chem Soc Rev. 44 (2015) 2732–2743. doi:10.1039/C4CS00257A.

[8] S. Zhang, J. Yang, R. Xu, F. Wang, W. Li, M. Ghufran, Y.-W. Zhang, Z. Yu, G. Zhang, Q. Qin, Y. Lu, Extraordinary Photoluminescence and Strong Temperature/Angle-Dependent Raman Responses in Few-Layer Phosphorene, ACS Nano. 8 (2014) 9590–9596. doi:10.1021/nn503893j.

[9] H. Liu, A.T. Neal, Z. Zhu, Z. Luo, X. Xu, D. Tománek, P.D. Ye, Phosphorene: An Unexplored 2D Semiconductor with a High Hole Mobility, ACS Nano. 8 (2014) 4033–4041. doi:10.1021/nn501226z.

[10] M. Buscema, D.J. Groenendijk, S.I. Blanter, G.A. Steele, H.S.J. van der Zant, A. Castellanos-Gomez, Fast and Broadband Photoresponse of Few-Layer Black Phosphorus Field-Effect Transistors, Nano Lett. 14 (2014) 3347–3352. doi:10.1021/nl5008085.

[11] J. Qiao, X. Kong, Z.-X. Hu, F. Yang, W. Ji, High-mobility transport anisotropy and linear dichroism in few-layer black phosphorus, Nat. Commun. 5 (2014) 4475. doi:10.1038/ncomms5475.

[12] A.S. Rodin, A. Carvalho, A.H. Castro Neto, Strain-Induced Gap Modification in Black

Phosphorus, Phys. Rev. Lett. 112 (2014) 176801. doi:10.1103/PhysRevLett.112.176801.
[13] X. Wang, A.M. Jones, K.L. Seyler, V. Tran, Y. Jia, H. Zhao, H. Wang, L. Yang, X. Xu, F. Xia, Highly anisotropic and robust excitons in monolayer black phosphorus, Nat. Nanotechnol. 10 (2015) 517–521. doi:10.1038/nnano.2015.71.
[14] N. Mao, J. Tang, L. Xie, J. Wu, B. Han, J. Lin, S. Deng, W. Ji, H. Xu, K. Liu, L. Tong, J. Zhang, Optical Anisotropy of Black Phosphorus in the Visible Regime, J. Am. Chem. Soc. 138 (2016) 300–305. doi:10.1021/jacs.5b10685.
[15] I. Kriegel, S. Toffanin, F. Scotognella, Black phosphorus-based one-dimensional photonic crystals and microcavities, Appl. Opt. 55 (2016) 9288. doi:10.1364/AO.55.009288.
[16] Q. Li, Q. Zhou, X. Niu, Y. Zhao, Q. Chen, J. Wang, Covalent Functionalization of Black Phosphorus from First-Principles, J. Phys. Chem. Lett. 7 (2016) 4540–4546. doi:10.1021/acs.jpclett.6b02192.
[17] C.R. Ryder, J.D. Wood, S.A. Wells, Y. Yang, D. Jariwala, T.J. Marks, G.C. Schatz, M.C. Hersam, Covalent functionalization and passivation of exfoliated black phosphorus via aryl diazonium chemistry, Nat. Chem. 8 (2016) 597–602. doi:10.1038/nchem.2505.
[18] X. Zhang, H. Xie, Z. Liu, C. Tan, Z. Luo, H. Li, J. Lin, L. Sun, W. Chen, Z. Xu, L. Xie, W. Huang, H. Zhang, Black Phosphorus Quantum Dots, Angew. Chem. Int. Ed. 54 (2015) 3653–3657. doi:10.1002/anie.201409400.
[19] Z. Sun, H. Xie, S. Tang, X.-F. Yu, Z. Guo, J. Shao, H. Zhang, H. Huang, H. Wang, P.K. Chu, Ultrasmall Black Phosphorus Quantum Dots: Synthesis and Use as Photothermal Agents, Angew. Chem. Int. Ed. 54 (2015) 11526–11530. doi:10.1002/anie.201506154.
[20] X. Niu, Y. Li, H. Shu, J. Wang, Anomalous Size Dependence of Optical Properties in Black Phosphorus Quantum Dots, J. Phys. Chem. Lett. 7 (2016) 370–375. doi:10.1021/acs.jpclett.5b02457.
[21] S. Zhou, N. Liu, J. Zhao, Phosphorus quantum dots as visible-light photocatalyst for water splitting, Comput. Mater. Sci. 130 (2017) 56–63. doi:10.1016/j.commatsci.2017.01.009.
[22] B. Rajbanshi, M. Kar, P. Sarkar, P. Sarkar, Phosphorene quantum dot-fullerene nanocomposites for solar energy conversion: An unexplored inorganic-organic nanohybrid with novel photovoltaic properties, Chem. Phys. Lett. 685 (2017) 16–22. doi:10.1016/j.cplett.2017.07.033.
[23] M.D. Hanwell, D.E. Curtis, D.C. Lonie, T. Vandermeersch, E. Zurek, G.R. Hutchison, Avogadro: an advanced semantic chemical editor, visualization, and analysis platform, J. Cheminformatics. 4 (2012) 17. doi:10.1186/1758-2946-4-17.
[24] F. Neese, The ORCA program system, Wiley Interdiscip. Rev. Comput. Mol. Sci. 2 (2012) 73–78. doi:10.1002/wcms.81.
[25] C. Lee, W. Yang, R.G. Parr, Development of the Colle-Salvetti correlation-energy formula into a functional of the electron density, Phys. Rev. B. 37 (1988) 785–789. doi:10.1103/PhysRevB.37.785.
[26] A. Schäfer, H. Horn, R. Ahlrichs, Fully optimized contracted Gaussian basis sets for atoms Li to Kr, J. Chem. Phys. 97 (1992) 2571–2577. doi:10.1063/1.463096.
[27] A. Schäfer, C. Huber, R. Ahlrichs, Fully optimized contracted Gaussian basis sets of triple zeta valence quality for atoms Li to Kr, J. Chem. Phys. 100 (1994) 5829–5835. doi:10.1063/1.467146.
[28] K. Eichkorn, F. Weigend, O. Treutler, R. Ahlrichs, Auxiliary basis sets for main row atoms and transition metals and their use to approximate Coulomb potentials, Theor. Chem. Acc. Theory Comput. Model. Theor. Chim. Acta. 97 (1997) 119–124. doi:10.1007/s002140050244.
[29] E.~F.~Valeev, A library for the evaluation of molecular integrals of many-body operators over Gaussian functions, (2014). http://libint.valeyev.net/.
[30] G. Malloci, C. Joblin, G. Mulas, On-line database of the spectral properties of polycyclic

aromatic hydrocarbons, Chem. Phys. 332 (2007) 353–359.
doi:10.1016/j.chemphys.2007.01.001.

**Supporting Information**

Covalent Functionalized Black Phosphorus Quantum Dots

Francesco Scotognella, Ilka Kriegel, Simone Sassolini

Optimized ground state geometries

P24H12

| | | | |
|---|---|---|---|
| P | -1.25844122423444 | 1.64868160769040 | -0.36955178534795 |
| P | -1.45697978275143 | -1.74462085704600 | -0.35531349670570 |
| P | -0.55003880248827 | 3.20700045700337 | 1.15425284762566 |
| P | -1.15449851658023 | -0.04030811735882 | 1.15767607306677 |
| H | -2.64385862948508 | 1.90062870365671 | -0.07605458650276 |
| P | -1.28169234165722 | -3.54840225938810 | 1.07315037992967 |
| H | -2.88717733255836 | -1.74102476064737 | -0.19007870107482 |
| H | 1.98336282887540 | 5.55656175746261 | 1.01370300520265 |
| P | 1.65335675425358 | 2.87783238274674 | 0.66813411936899 |
| H | -0.61987648435606 | 4.26984414743690 | 0.18772228520137 |
| P | 1.09721378809188 | -0.42505926215106 | 1.00686611790898 |
| P | 1.00190549879661 | -3.63286191768057 | 1.33356739890167 |
| H | -1.26283195096859 | -4.45566147139721 | -0.04358000787750 |
| P | 0.94927894762320 | -5.25611122058368 | 2.94455038461125 |
| P | 0.85169197000965 | -1.86668707087282 | 2.76237528334047 |
| P | 2.32310832144666 | 1.18501966130110 | 1.98278228009026 |
| P | 2.77148281623705 | 4.55072079731896 | 1.66818894617918 |
| P | 3.16427196814311 | -5.03053403584250 | 3.49571919023232 |
| P | 2.93745965151166 | -1.67898994744052 | 3.62754580170545 |
| P | 4.88496143098831 | -3.12970332350382 | 5.78709692540034 |
| P | 4.19913691204377 | 0.29716144644799 | 5.99010296905900 |
| H | 4.78608775343987 | -4.17567047676134 | 6.77056483641773 |
| H | 1.11381829944068 | 2.69094037679592 | 6.73516383566872 |
| P | 2.38994274710696 | 3.34273607793862 | 6.77008654449684 |
| H | 2.80763369926281 | 5.78331068485624 | 4.10621160761254 |
| P | 1.38658320484789 | 1.56400427635278 | 4.02327734403996 |
| P | 4.31223635298430 | -1.55236002292512 | 7.35290207875345 |
| P | 2.73807222672689 | -3.43753979059691 | 5.07341004596111 |
| P | 1.79361743028922 | 4.83849968948866 | 3.72127820390722 |
| P | 3.48835139209408 | 1.62253950257778 | 7.72632630202290 |
| H | 5.69018760913282 | -1.29321585918562 | 7.67403703414941 |
| P | 2.11322591434437 | -0.26169754702285 | 5.23900564571347 |
| H | 4.72766545650282 | 2.34081887899211 | 7.77757279822603 |
| P | 2.96279252596164 | 3.09142520534990 | 4.60177784548323 |
| H | 3.07901632066051 | -6.08759456662986 | 4.46684400623413 |
| H | 1.23193324426392 | -6.28898314638261 | 1.98448644099796 |

BenzP24H11

| | | | |
|---|---|---|---|
| P | -0.87331358690269 | 0.38319457373831 | -1.34967080215516 |
| P | -1.93979537938937 | -2.71693820960271 | -0.43651662887038 |
| P | 0.12730953555329 | 2.11208573870841 | -0.22612586276455 |
| P | -1.28302064623275 | -0.76235887037324 | 0.57912265985369 |
| H | -2.16249541345565 | 1.01935217893881 | -1.30800828410285 |
| P | -2.30568017649982 | -4.03362867559102 | 1.42460406062172 |
| H | -3.32962864726936 | -2.34737151294614 | -0.45443181631990 |
| H | 3.17286531961124 | 3.67450230207871 | -0.67210722849183 |
| P | 2.19602619614150 | 1.15860306426517 | -0.33171866173583 |
| H | 0.38424025908092 | 2.83620054736111 | -1.44217239928106 |
| P | 0.79827566258064 | -1.68655394153333 | 0.80487107126439 |
| P | -0.12675709400978 | -4.54954006319789 | 1.96100230638558 |
| C | -2.74808146000294 | -5.58431906930742 | 0.49975744889209 |
| P | -0.66153307513253 | -5.59865035480747 | 3.91947264219815 |
| P | 0.10865051014258 | -2.46390016123203 | 2.83842878870879 |
| P | 2.34581573873055 | -0.19296304499320 | 1.45452914506069 |
| P | 3.65159952375697 | 2.74473524828315 | 0.31159878633184 |
| P | 1.50913615920737 | -5.73424229429054 | 4.64640818577091 |
| P | 2.13084111583027 | -2.52436401352126 | 3.86026312562586 |
| P | 3.53733439987998 | -3.70609300727690 | 6.54225704151956 |
| P | 3.73949998818636 | -0.30121101495042 | 5.74838190384896 |
| H | 3.12659851134765 | -4.37624889708507 | 7.74769970525184 |
| H | 1.33139525396714 | 2.85361200577585 | 5.46171383635082 |
| P | 2.72966120910083 | 3.17082661323093 | 5.47145504949151 |
| H | 3.88864668829356 | 4.57605668724951 | 2.32522023316752 |
| P | 1.43702823651891 | 0.95957228462808 | 3.19601618108670 |
| P | 3.31078820675840 | -1.65960798501037 | 7.55551232532372 |
| P | 1.42166911522366 | -3.70325642064886 | 5.68383449238806 |
| P | 2.68459120150230 | 3.82542149579167 | 2.08778355214957 |
| P | 3.30088671377179 | 1.59114270833751 | 6.97402896185861 |
| H | 4.69225691850496 | -1.64565121422375 | 7.95580990435825 |
| P | 1.61879585579087 | -0.55568855045257 | 4.93134971919817 |
| H | 4.67730499220246 | 1.98976562937047 | 6.98718660509870 |
| P | 3.32023401942470 | 2.18031898065069 | 3.53103263308570 |
| H | 1.11827208125203 | -6.42111283126891 | 5.84784372632298 |
| H | -0.59937003224738 | -6.89421653018261 | 3.30035414608905 |
| C | -3.95268470084088 | -6.20913777786222 | 0.87001758322612 |
| C | -1.96685219532535 | -6.16369163211226 | -0.51857124002309 |
| C | -4.36850422195641 | -7.38635933540155 | 0.23860543062378 |
| C | -3.58393921897655 | -7.95366126449834 | -0.76942242252585 |
| C | -2.38306449767076 | -7.34007861358274 | -1.14615517048803 |
| H | -1.02736174684499 | -5.69330730493416 | -0.82314409951606 |
| H | -4.56918918504394 | -5.77017842898823 | 1.66158847742448 |
| H | -5.30859193449227 | -7.86002135377949 | 0.53737272521007 |
| H | -3.90666429425698 | -8.87566735500937 | -1.26248230672850 |
| H | -1.76579590581052 | -7.78067032974432 | -1.93549553078478 |

2BenzP24H10

| | | | |
|---|---|---|---|
| P | -0.71781911267840 | 0.40386851841526 | -1.32707841723897 |
| P | -1.99326351958660 | -2.59319394620688 | -0.34912168867679 |
| P | 0.52242564947059 | 2.01048336624176 | -0.26516777223081 |
| P | -1.05532233672394 | -0.73345182531599 | 0.61873134536185 |
| H | -1.93840625609392 | 1.14811003071647 | -1.16751622092056 |
| P | -2.33195705849676 | -3.88007194096636 | 1.53550909035013 |
| H | -3.33250814842466 | -2.07830757575028 | -0.24810925078277 |
| H | 3.59459016675770 | 3.34538398281864 | -0.94416842360920 |
| P | 2.48476603012194 | 0.89173239918814 | -0.57769198059156 |
| H | 0.73721918192649 | 2.74346426773663 | -1.48444834810136 |
| P | 0.94191571521290 | -1.85494852826430 | 0.64389035010482 |
| P | -0.18034878162313 | -4.62431566133927 | 1.86775827189038 |
| C | -3.01657084982388 | -5.36925965251436 | 0.65804386226873 |
| P | -0.65125189373546 | -5.63980175554459 | 3.85972376439982 |
| P | 0.36049477849706 | -2.58785388704302 | 2.73000945712306 |
| P | 2.68891580586141 | -0.52881111807889 | 1.15291071956842 |
| P | 4.09532934491781 | 2.36081473236020 | -0.02602544112070 |
| P | 1.54581861457940 | -6.03475550931883 | 4.38160329044339 |
| P | 2.44944584693325 | -2.90098722514870 | 3.55170112032446 |
| P | 3.95942568320636 | -4.29953205159830 | 6.08589659367589 |
| P | 4.53447936795232 | -0.98030467134139 | 5.20921441631555 |
| H | 3.57750610017256 | -4.91086681726663 | 7.33133000337159 |
| H | -2.50154120008319 | -7.62959616710341 | -1.86095243213982 |
| P | 3.55907168786962 | 2.48125509984955 | 5.24403829063813 |
| H | 4.60851119060796 | 4.13698991907937 | 1.97399581337710 |
| P | 2.05859576020266 | 0.64322711887749 | 2.99732622452943 |
| P | 4.11604855293538 | -2.23938922420250 | 7.08472328901966 |
| P | 1.78939162399807 | -4.02200157809094 | 5.42983992553360 |
| P | 3.35357740625351 | 3.44400854233122 | 1.85502045952891 |
| P | 4.76845018117249 | 0.95095654524349 | 6.42514406353127 |
| H | 5.52009342087540 | -2.41011737743417 | 7.34535050653843 |
| P | 2.30761639992881 | -0.90557709575394 | 4.68853589027453 |
| H | 6.03524668990045 | 1.22549414286427 | 5.81286777710918 |
| P | 4.06390778450724 | 1.71377559117278 | 3.15888920303766 |
| H | 1.19109284896391 | -6.68854421228476 | 5.61261333819349 |
| H | -0.79780136522552 | -6.92214076426504 | 3.22717150864706 |
| C | -4.23834968386624 | -5.87178472603071 | 1.14179604495161 |
| C | -2.39936412086811 | -6.01523376255969 | -0.43053706930523 |
| C | -4.83152443341705 | -6.99376077282847 | 0.55219170683047 |
| C | -4.20881507930685 | -7.62776004011772 | -0.52671601094832 |
| C | -2.99226034926538 | -7.13653209154357 | -1.01615076789514 |
| H | -1.44977823043600 | -5.64039247055584 | -0.82259671530202 |
| H | -4.72801715458755 | -5.38081111556889 | 1.98895996812478 |
| H | -5.78268457908560 | -7.37172886505183 | 0.93864537152101 |
| H | -4.67004005086593 | -8.50652691975481 | -0.98756977987161 |
| C | 4.71842913814298 | 3.92261193823676 | 5.40376288241029 |
| C | 6.01634840169332 | 3.97592616501499 | 4.86184569463895 |
| C | 4.23019568205530 | 5.04029951778492 | 6.10442110306914 |
| C | 5.02364697644475 | 6.18121657092925 | 6.26733907436722 |
| C | 6.80431137385629 | 5.11905469967759 | 5.01828372414106 |
| C | 6.31065210994194 | 6.22329931610575 | 5.72397099029625 |
| H | 6.41946160128694 | 3.11829760196161 | 4.31349710374509 |
| H | 3.21977872552367 | 5.01783437905074 | 6.52397008357904 |
| H | 4.63049246053675 | 7.04140880004264 | 6.81810898345184 |
| H | 7.81031556188056 | 5.14608997891553 | 4.58888199111983 |
| H | 6.93095634000636 | 7.11645612422902 | 5.84823702133176 |

4BenzP24H8

| | | | |
|---|---|---|---|
| P | -0.88656510501728 | 0.32648092956823 | -1.28907468587580 |
| P | -2.02815815861266 | -2.69651738425221 | -0.24352123127212 |
| P | 0.32193743741290 | 2.00726058247873 | -0.31188361382561 |
| P | -1.11959922023382 | -0.79693344915250 | 0.67655487336890 |
| H | -2.12252459592770 | 1.03814196417221 | -1.10854175805534 |
| P | -2.24504215567214 | -3.98121124570488 | 1.65934518706473 |
| H | -3.38053730876708 | -2.23009310775978 | -0.09245183622317 |
| H | 3.33864940160128 | 3.43654134954275 | -1.08339464418351 |
| P | 2.32117857105779 | 0.95545637341552 | -0.66494762594298 |
| C | 0.39266313899919 | 3.15806244240277 | -1.76959953044359 |
| P | 0.91564289930095 | -1.84212504459761 | 0.63037223664420 |
| P | -0.05561895958978 | -4.64437281616157 | 1.90735431509735 |
| C | -2.90854410537860 | -5.49992319012781 | 0.81597465874342 |
| P | -0.40501810816601 | -5.67408481760630 | 3.91638724194024 |
| P | 0.44924102625784 | -2.58700998850869 | 2.74144236730476 |
| P | 2.62669884069648 | -0.44712004998472 | 1.06549348434794 |
| P | 3.90286191353589 | 2.47546231991014 | -0.17983078093498 |
| P | 1.82363344895185 | -5.98787415082352 | 4.34557995492729 |
| P | 2.58220280813015 | -2.82217159703864 | 3.47194304347730 |
| P | 4.24550858788142 | -4.17031220546985 | 5.94731133303489 |
| P | 4.65502625926806 | -0.80651442549812 | 5.04448071048551 |
| C | 4.04341573083149 | -5.14949321757620 | 7.51605710959713 |
| H | -2.40944468432624 | -7.75458741668333 | -1.71159678290953 |
| P | 3.55388705537409 | 2.62352704624155 | 5.09976555892175 |
| H | 4.42352288027885 | 4.27621919375360 | 1.80142739736595 |
| P | 2.03123305364575 | 0.70607166302262 | 2.93509810631068 |
| P | 4.37217343753446 | -2.09746538456820 | 6.92307623350294 |
| P | 2.04221711795784 | -3.96500686438140 | 5.37420248532795 |
| P | 3.18976193151784 | 3.54101754464024 | 1.71981937884101 |
| P | 4.82712969797959 | 1.12698823596925 | 6.26691062342018 |
| H | 5.78851581617149 | -2.22976287961498 | 7.13011865348159 |
| P | 2.40976240768662 | -0.83264937501397 | 4.61262331286848 |
| H | 6.07694832625783 | 1.44891388537104 | 5.64262193585390 |
| P | 4.00278643211413 | 1.84558288755703 | 3.00703325983523 |
| H | 1.54713860051418 | -6.64632357972554 | 5.59286074843298 |
| H | -0.53184333690331 | -6.96016746582558 | 3.28656529173486 |
| C | -4.09287607555952 | -6.04259470040822 | 1.34725588069183 |
| C | -2.31048171792963 | -6.13008352689642 | -0.29239253956127 |
| C | -4.66814916297206 | -7.18764700384481 | 0.78490557160605 |
| C | -4.06464942204086 | -7.80535712603675 | -0.31419834563048 |
| C | -2.88516744009379 | -7.27453614526239 | -0.85088345733853 |
| H | -1.38982065669669 | -5.72445739156222 | -0.72123645045966 |
| H | -4.56708830178432 | -5.56460336487376 | 2.21044880357909 |
| H | -5.59044206930437 | -7.59645776465824 | 1.20843714362169 |
| H | -4.51178505831834 | -8.70203322930856 | -0.75393377364991 |
| C | 4.69683204111390 | 4.08302740358442 | 5.21235992265223 |
| C | 5.97447616976994 | 4.15159067206164 | 4.62554960103262 |
| C | 4.21855946651067 | 5.19654858192416 | 5.92635641975555 |
| C | 5.00266759018182 | 6.34797132243671 | 6.05952300657800 |
| C | 6.75286800977178 | 5.30474638171524 | 4.75259248967335 |
| C | 6.26976951547288 | 6.40451155873051 | 5.47268501106570 |
| H | 6.36741269876110 | 3.29736548896266 | 4.06490115871836 |
| H | 3.22300943867600 | 5.16189763409545 | 6.37943429174095 |
| H | 4.61785141533636 | 7.20494749393610 | 6.62110992787768 |
| H | 7.74333379366818 | 5.34346920040537 | 4.28914016427311 |
| H | 6.88271448977318 | 7.30565904029584 | 5.57428477046543 |
| C | 0.10225733262908 | 4.51001525668652 | -1.50905134291129 |
| C | 0.11853929346399 | 5.45439821614261 | -2.54110226328489 |
| C | 0.42952773392634 | 5.06039901977053 | -3.84582102492584 |
| C | 0.72378282572982 | 3.71858640772091 | -4.11594157837072 |
| C | 0.70492316687090 | 2.77273436456306 | -3.08781605660959 |
| H | -0.14171232011272 | 4.82398472762867 | -0.48899705440658 |
| H | -0.11533153901155 | 6.50121314402909 | -2.32285617571986 |
| H | 0.44255537360221 | 5.79815157847243 | -4.65434252566628 |
| H | 0.96982621686569 | 3.40490182502058 | -5.13534836746976 |
| H | 0.93557941054563 | 1.72720985959933 | -3.31327619698596 |
| C | 3.15162880931061 | -4.82642648011428 | 8.55696049582758 |
| C | 3.07406319721128 | -5.63024626654462 | 9.69738892399928 |
| C | 3.88326555108712 | -6.76686759544926 | 9.81621989607453 |
| C | 4.77069968232497 | -7.09876116371870 | 8.78836599121552 |
| C | 4.85035763815700 | -6.29464319841464 | 7.64566511758402 |
| H | 2.51596986624666 | -3.93966223316559 | 8.47765566244351 |
| H | 2.37748170813610 | -5.36652239986345 | 10.49972091046673 |
| H | 3.82043147727841 | -7.39345431831305 | 10.71166546544237 |
| H | 5.40423862885988 | -7.98734163753818 | 8.87303417617791 |
| H | 5.54644014017900 | -6.55924039377893 | 6.84255933816301 |

AnthP24H11

| | | | |
|---|---|---|---|
| P | -0.83238189846001 | 0.34285849165933 | -1.36890200084029 |
| P | -1.86066842139192 | -2.77598375480551 | -0.46312438595531 |
| P | 0.13241349880384 | 2.08484060994276 | -0.23465709118120 |
| P | -1.24637280403353 | -0.80840924048620 | 0.55514454820116 |
| H | -2.12996752130398 | 0.96258078712104 | -1.34034866114353 |
| P | -2.23500972468109 | -4.08762398200229 | 1.39941146512152 |
| H | -3.25535232703048 | -2.42675047492118 | -0.50604929601751 |
| H | 3.15603988217262 | 3.69126518364837 | -0.64855072917598 |
| P | 2.21532137370473 | 1.16122952176794 | -0.31989123559637 |
| H | 0.39123981486305 | 2.81230999341893 | -1.44845201555178 |
| P | 0.84564793369287 | -1.70377770098631 | 0.80146755391529 |
| P | -0.05679327550094 | -4.58291156686597 | 1.95024218699566 |
| C | -2.67184296096102 | -5.63767656658830 | 0.46982673914424 |
| P | -0.59122584246907 | -5.63557473500811 | 3.90666319928356 |
| P | 0.14821662495845 | -2.49282114488049 | 2.82788029850677 |
| P | 2.36642887715571 | -0.19043834617258 | 1.46652320297111 |
| P | 3.64019230909363 | 2.76847222121084 | 0.33910935894216 |
| P | 1.57638265688026 | -5.74492302946319 | 4.64702473967699 |
| P | 2.16205065224888 | -2.52700940256812 | 3.86685576492609 |
| P | 3.56430458094988 | -3.69014315115735 | 6.55839427985580 |
| P | 3.72159183244030 | -0.28119685055144 | 5.77135657696756 |
| H | 3.15454540007656 | -4.36764834783899 | 7.76016526056364 |
| H | 1.26671999235644 | 2.83376764018585 | 5.46470141273993 |
| P | 2.65948136846826 | 3.17408635947527 | 5.48915695114371 |
| H | 3.83031522115874 | 4.60131390938255 | 2.35584024861962 |
| P | 1.42526638037153 | 0.94794190181193 | 3.19919740390447 |
| P | 3.30063014461076 | -1.64856694355332 | 7.57312891747364 |
| P | 1.45521931389486 | -3.71621337400940 | 5.68509769998439 |
| P | 2.63961296478698 | 3.83327326736032 | 2.10620297248164 |
| P | 3.24047371943447 | 1.60148113368959 | 6.99560258701999 |
| H | 4.67927751873033 | -1.61501587884774 | 7.98214873411233 |
| P | 1.61237306472961 | -0.56726469591614 | 4.93472833337128 |
| H | 4.61027453517352 | 2.02188758296959 | 7.02338765746365 |
| P | 3.28635480376403 | 2.19634750155601 | 3.55383053142741 |
| H | 1.18559573189947 | -6.43776009009917 | 5.84511018900381 |
| H | -0.50948016764112 | -6.93103191237685 | 3.28953234685718 |
| C | -3.93682300153417 | -6.23350082921314 | 0.79864844439197 |
| C | -1.87467065261999 | -6.23702554790042 | -0.48644345310682 |
| C | -4.36009270763686 | -7.38196468202615 | 0.17784922602544 |
| C | -3.55498456446158 | -8.02720631705471 | -0.81420927133842 |
| C | -2.27841163171201 | -7.43678681580012 | -1.15132772294390 |
| C | -3.95635613129337 | -9.20326638768098 | -1.46451713380029 |
| C | -1.47235979638630 | -8.05637676497792 | -2.12054461683865 |
| C | -3.15121745707931 | -9.82377584184400 | -2.43277806030513 |

```
C   -1.87313539090768     -9.23322191802295     -2.77012013200768
C   -3.54859549026338    -11.02612689492398     -3.10160472002415
C   -2.73833858225977    -11.61117792558060     -4.04321849969966
C   -1.47802907818641    -11.02792189656947     -4.37727456209928
C   -1.05986962237545     -9.87546600623985     -3.75960610186553
H   -4.56605405018115     -5.75895118210456      1.55853277671713
H   -0.90813466405275     -5.80088574585937     -0.75731734765623
H   -5.32723483343768     -7.82328190163176      0.43824749049367
H   -4.92329105724658     -9.64905442669396     -1.20986888963124
H   -0.50600116608709     -7.60922193818183     -2.37471944472028
H   -4.51491416832318    -11.47171027397884     -2.84551338591448
H   -3.05437598025688    -12.53143182165318     -4.54363144372767
H   -0.84623507532110    -11.50772689011906     -5.13118678755637
H   -0.09415015132394     -9.42783290804482     -4.01405210960505
```

Transitions for the different studied materials (for the first 16 excited singlet states).

P24H12

ABSORPTION SPECTRUM VIA TRANSITION ELECTRIC DIPOLE MOMENTS

| State | Energy (cm-1) | Wavelength (nm) | fosc | T2 (au**2) | TX (au) | TY (au) | TZ (au) |
|---|---|---|---|---|---|---|---|
| 1 | 24189.6 | 413.4 | 0.008844182 | 0.12037 | -0.15322 | 0.17605 | -0.25670 |
| 2 | 24577.6 | 406.9 | 0.000555437 | 0.00744 | 0.05453 | 0.00660 | 0.06651 |
| 3 | 25941.1 | 385.5 | 0.010396761 | 0.13194 | -0.20598 | 0.04384 | -0.29596 |
| 4 | 26735.3 | 374.0 | 0.010135038 | 0.12480 | 0.17895 | 0.07878 | 0.29423 |
| 5 | 27119.3 | 368.7 | 0.012572244 | 0.15262 | 0.09612 | 0.37529 | 0.05037 |
| 6 | 27402.1 | 364.9 | 0.015268270 | 0.18343 | 0.19599 | -0.37141 | 0.08412 |
| 7 | 27510.5 | 363.5 | 0.007937467 | 0.09499 | 0.11492 | -0.26156 | 0.11560 |
| 8 | 27853.9 | 359.0 | 0.014137222 | 0.16709 | -0.25426 | -0.11803 | -0.29751 |
| 9 | 28262.2 | 353.8 | 0.002582742 | 0.03009 | -0.13632 | 0.07950 | -0.07198 |
| 10 | 28353.9 | 352.7 | 0.026658589 | 0.30953 | 0.20029 | 0.42241 | 0.30163 |
| 11 | 29139.8 | 343.2 | 0.006374769 | 0.07202 | -0.13041 | 0.20896 | -0.10654 |
| 12 | 29380.2 | 340.4 | 0.008897367 | 0.09970 | 0.05356 | 0.09422 | 0.29657 |
| 13 | 29445.2 | 339.6 | 0.010362280 | 0.11586 | 0.21081 | -0.12999 | 0.23349 |
| 14 | 29472.5 | 339.3 | 0.013852929 | 0.15474 | -0.00430 | 0.39267 | -0.02302 |
| 15 | 29574.6 | 338.1 | 0.011103455 | 0.12360 | -0.23734 | -0.03826 | -0.25653 |
| 16 | 29750.1 | 336.1 | 0.014696821 | 0.16263 | 0.26992 | -0.08584 | 0.28706 |

BenzP24H11

ABSORPTION SPECTRUM VIA TRANSITION ELECTRIC DIPOLE MOMENTS

| State | Energy (cm-1) | Wavelength (nm) | fosc | T2 (au**2) | TX (au) | TY (au) | TZ (au) |
|---|---|---|---|---|---|---|---|
| 1 | 24133.6 | 414.4 | 0.008928501 | 0.12180 | -0.09606 | 0.12331 | -0.31203 |
| 2 | 24569.7 | 407.0 | 0.001559209 | 0.02089 | -0.07994 | -0.01233 | -0.11979 |
| 3 | 25872.8 | 386.5 | 0.009990405 | 0.12712 | -0.17901 | -0.02169 | -0.30758 |
| 4 | 26497.5 | 377.4 | 0.012159820 | 0.15108 | 0.20102 | 0.11110 | 0.31357 |
| 5 | 26962.5 | 370.9 | 0.017826109 | 0.21766 | 0.24184 | 0.39889 | 0.00761 |
| 6 | 27250.7 | 367.0 | 0.018568051 | 0.22432 | 0.14894 | -0.34259 | 0.29115 |
| 7 | 27540.0 | 363.1 | 0.002106054 | 0.02518 | 0.00009 | -0.11707 | 0.10710 |
| 8 | 27763.5 | 360.2 | 0.017058304 | 0.20227 | -0.28090 | -0.21365 | -0.27878 |
| 9 | 28123.0 | 355.6 | 0.003620036 | 0.04238 | 0.00257 | -0.20524 | 0.01571 |
| 10 | 28197.2 | 354.6 | 0.017890988 | 0.20888 | 0.27731 | 0.29933 | 0.20587 |
| 11 | 28650.8 | 349.0 | 0.009013930 | 0.10357 | -0.01968 | -0.32088 | 0.01489 |
| 12 | 29049.9 | 344.2 | 0.005386925 | 0.06105 | -0.15903 | 0.09539 | -0.16327 |
| 13 | 29235.9 | 342.0 | 0.017257124 | 0.19432 | -0.16756 | -0.03930 | -0.40584 |
| 14 | 29344.7 | 340.8 | 0.003474591 | 0.03898 | 0.07795 | -0.11354 | 0.14147 |
| 15 | 29450.2 | 339.6 | 0.012513309 | 0.13988 | 0.00696 | 0.30234 | -0.22005 |
| 16 | 29611.1 | 337.7 | 0.015812125 | 0.17580 | 0.28444 | 0.19807 | 0.23591 |

2BenzP24H10

ABSORPTION SPECTRUM VIA TRANSITION ELECTRIC DIPOLE MOMENTS

| State | Energy (cm-1) | Wavelength (nm) | fosc | T2 (au**2) | TX (au) | TY (au) | TZ (au) |
|---|---|---|---|---|---|---|---|
| 1 | 24002.4 | 416.6 | 0.001835638 | 0.02518 | 0.03283 | -0.06656 | 0.14025 |
| 2 | 24426.7 | 409.4 | 0.004674415 | 0.06300 | 0.09104 | -0.06400 | 0.22498 |
| 3 | 25696.7 | 389.2 | 0.015994956 | 0.20492 | 0.25762 | -0.05273 | 0.36847 |
| 4 | 26484.0 | 377.6 | 0.018714837 | 0.23264 | -0.36220 | -0.22662 | -0.22380 |
| 5 | 26676.8 | 374.9 | 0.020750223 | 0.25607 | 0.19567 | 0.46476 | -0.04225 |
| 6 | 26947.6 | 371.1 | 0.000479482 | 0.00586 | -0.06250 | -0.01151 | 0.04265 |
| 7 | 27403.7 | 364.9 | 0.018202093 | 0.21867 | 0.05618 | 0.44183 | -0.14247 |
| 8 | 27688.0 | 361.2 | 0.014993849 | 0.17828 | -0.29030 | 0.09621 | -0.29111 |
| 9 | 27942.1 | 357.9 | 0.007736650 | 0.09115 | 0.17381 | 0.14656 | 0.19866 |
| 10 | 28009.3 | 357.0 | 0.005613814 | 0.06598 | 0.09107 | 0.23588 | -0.04525 |
| 11 | 28701.6 | 348.4 | 0.003678991 | 0.04220 | -0.13606 | -0.06852 | -0.13781 |
| 12 | 28846.7 | 346.7 | 0.000218094 | 0.00249 | -0.03301 | -0.02556 | 0.02731 |
| 13 | 28869.7 | 346.4 | 0.011073501 | 0.12628 | 0.10152 | 0.34040 | 0.01001 |
| 14 | 29165.3 | 342.9 | 0.013554321 | 0.15300 | 0.15450 | 0.35911 | -0.01301 |
| 15 | 29240.3 | 342.0 | 0.013115726 | 0.14767 | -0.32048 | -0.20377 | -0.05861 |
| 16 | 29293.6 | 341.4 | 0.021638625 | 0.24318 | 0.15537 | -0.23026 | 0.40746 |

4BenzP24H8

ABSORPTION SPECTRUM VIA TRANSITION ELECTRIC DIPOLE MOMENTS

| State | Energy (cm-1) | Wavelength (nm) | fosc | T2 (au**2) | TX (au) | TY (au) | TZ (au) |
|---|---|---|---|---|---|---|---|
| 1 | 23767.9 | 420.7 | 0.003086037 | 0.04275 | -0.06458 | 0.08283 | -0.17808 |
| 2 | 24233.9 | 412.6 | 0.002904170 | 0.03945 | -0.06345 | 0.06567 | -0.17639 |
| 3 | 25216.8 | 396.6 | 0.037589923 | 0.49075 | -0.35776 | 0.19693 | -0.56919 |
| 4 | 26397.5 | 378.8 | 0.018661639 | 0.23274 | -0.35091 | -0.27232 | -0.18826 |
| 5 | 26552.1 | 376.6 | 0.013766949 | 0.17069 | -0.13452 | -0.38642 | 0.05724 |
| 6 | 26891.4 | 371.9 | 0.000601173 | 0.00736 | 0.01209 | -0.00894 | -0.08446 |
| 7 | 27181.2 | 367.9 | 0.023559726 | 0.28535 | 0.06647 | 0.50313 | -0.16672 |
| 8 | 27405.5 | 364.9 | 0.017416245 | 0.20921 | 0.29408 | -0.16224 | 0.31050 |
| 9 | 27643.9 | 361.7 | 0.012333817 | 0.14688 | -0.24478 | -0.20725 | -0.20979 |
| 10 | 27680.4 | 361.3 | 0.006632152 | 0.07888 | 0.04984 | -0.17564 | 0.21341 |
| 11 | 28241.3 | 354.1 | 0.011688276 | 0.13625 | 0.02614 | -0.32717 | 0.16890 |
| 12 | 28372.4 | 352.5 | 0.027954923 | 0.32437 | -0.39028 | -0.32391 | -0.25910 |
| 13 | 28627.5 | 349.3 | 0.002937282 | 0.03378 | -0.03767 | 0.10968 | -0.14258 |
| 14 | 28740.0 | 347.9 | 0.003314120 | 0.03796 | 0.00559 | 0.17048 | -0.09417 |
| 15 | 28901.9 | 346.0 | 0.001088376 | 0.01240 | -0.06526 | 0.02415 | -0.08692 |
| 16 | 28990.8 | 344.9 | 0.003311519 | 0.03760 | 0.11703 | -0.13287 | 0.07908 |

AnthP24H11

ABSORPTION SPECTRUM VIA TRANSITION ELECTRIC DIPOLE MOMENTS

| State | Energy (cm-1) | Wavelength (nm) | fosc | T2 (au**2) | TX (au) | TY (au) | TZ (au) |
|---|---|---|---|---|---|---|---|
| 1 | 24099.0 | 415.0 | 0.012244203 | 0.16727 | 0.12751 | -0.09251 | 0.37742 |
| 2 | 24539.6 | 407.5 | 0.002598232 | 0.03486 | -0.10152 | -0.04281 | -0.15073 |
| 3 | 25105.6 | 398.3 | 0.007063933 | 0.09263 | 0.12554 | 0.01060 | 0.27705 |
| 4 | 25596.1 | 390.7 | 0.017337269 | 0.22299 | -0.40251 | 0.09303 | 0.22874 |
| 5 | 25660.6 | 389.7 | 0.024346285 | 0.31235 | 0.37004 | -0.12594 | -0.39944 |
| 6 | 26187.8 | 381.9 | 0.011467305 | 0.14416 | 0.15813 | 0.16051 | 0.30560 |
| 7 | 26453.2 | 378.0 | 0.007502749 | 0.09337 | -0.19677 | -0.18191 | 0.14683 |
| 8 | 26623.4 | 375.6 | 0.008326381 | 0.10296 | -0.03509 | -0.20676 | -0.24285 |
| 9 | 26850.6 | 372.4 | 0.002104832 | 0.02581 | -0.05702 | 0.14440 | 0.04128 |
| 10 | 26907.2 | 371.6 | 0.004999660 | 0.06117 | -0.10685 | 0.14559 | 0.16899 |
| 11 | 27131.7 | 368.6 | 0.030591079 | 0.37119 | 0.25348 | 0.55000 | 0.06664 |
| 12 | 27469.2 | 364.0 | 0.015471347 | 0.18542 | -0.26943 | 0.20534 | -0.26583 |
| 13 | 27645.3 | 361.7 | 0.011229632 | 0.13373 | 0.14744 | -0.22824 | 0.24473 |
| 14 | 27863.7 | 358.9 | 0.015905676 | 0.18793 | -0.15312 | -0.33920 | -0.22232 |
| 15 | 28143.7 | 355.3 | 0.014200966 | 0.16612 | 0.19126 | 0.31979 | 0.16515 |
| 16 | 28451.1 | 351.5 | 0.021532932 | 0.24916 | 0.35805 | 0.32008 | 0.13607 |

Anhtracene

ABSORPTION SPECTRUM VIA TRANSITION ELECTRIC DIPOLE MOMENTS

| State | Energy (cm-1) | Wavelength (nm) | fosc | T2 (au**2) | TX (au) | TY (au) | TZ (au) |
|---|---|---|---|---|---|---|---|
| 1 | 28078.0 | 356.2 | 0.089958467 | 1.05475 | -0.00000 | -0.00000 | -1.02701 |
| 2 | 31480.7 | 317.7 | 0.000535253 | 0.00560 | 0.00000 | -0.07482 | -0.00000 |
| 3 | 36856.9 | 271.3 | 0.000000000 | 0.00000 | -0.00000 | -0.00000 | 0.00000 |
| 4 | 41480.0 | 241.1 | 0.000000000 | 0.00000 | -0.00000 | 0.00000 | -0.00000 |
| 5 | 44001.0 | 227.3 | 0.000016549 | 0.00012 | -0.00000 | 0.00000 | -0.01113 |
| 6 | 45043.6 | 222.0 | 0.000000000 | 0.00000 | 0.00000 | -0.00000 | 0.00000 |
| 7 | 46334.2 | 215.8 | 3.077748217 | 21.86789 | -0.00000 | -4.67631 | -0.00000 |
| 8 | 48161.3 | 207.6 | 0.000000000 | 0.00000 | 0.00000 | 0.00000 | -0.00000 |
| 9 | 48609.9 | 205.7 | 0.064125287 | 0.43429 | -0.00000 | 0.00000 | 0.65901 |
| 10 | 49185.1 | 203.3 | 0.000051988 | 0.00035 | 0.01865 | -0.00000 | -0.00000 |
| 11 | 50976.0 | 196.2 | 0.000000000 | 0.00000 | -0.00000 | 0.00000 | 0.00000 |
| 12 | 51335.9 | 194.8 | 0.000000000 | 0.00000 | 0.00000 | 0.00000 | -0.00000 |
| 13 | 51706.0 | 193.4 | 0.080898589 | 0.51508 | 0.00000 | 0.00000 | -0.71769 |
| 14 | 52170.7 | 191.7 | 0.000000000 | 0.00000 | -0.00000 | 0.00000 | -0.00000 |
| 15 | 53216.2 | 187.9 | 0.000000000 | 0.00000 | -0.00000 | -0.00000 | -0.00000 |
| 16 | 53421.8 | 187.2 | 0.004739229 | 0.02921 | -0.00000 | -0.17090 | 0.00000 |

Orbitals of P24H12 and 4BenzP24H8

**- P24H12**

States involved in the lowest transistion at 2.999 eV (413.4 nm).
Only excitations with weight larger than 0.01 are reported.

185a -> 186a : 0.948613
185a -> 187a : 0.026173

Orbital 185: HOMO

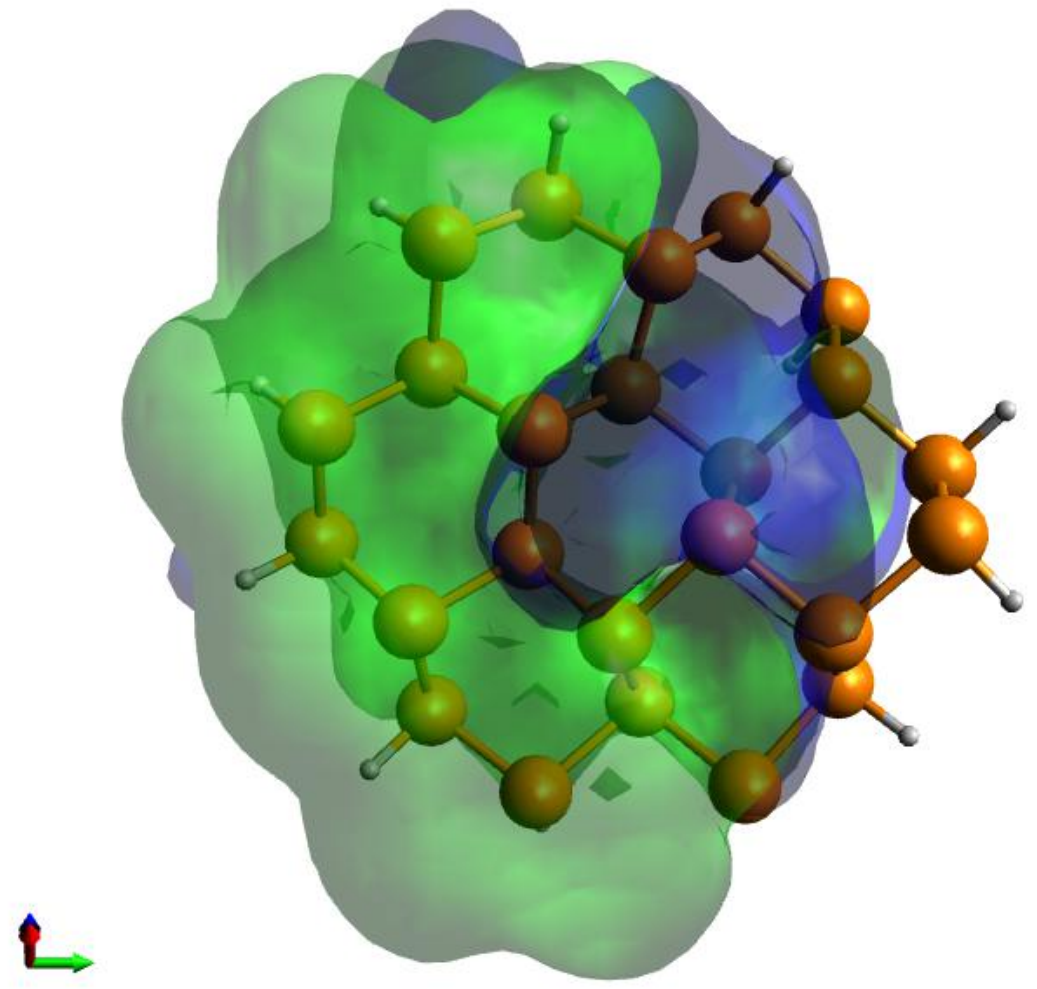

Orbital 186: LUMO

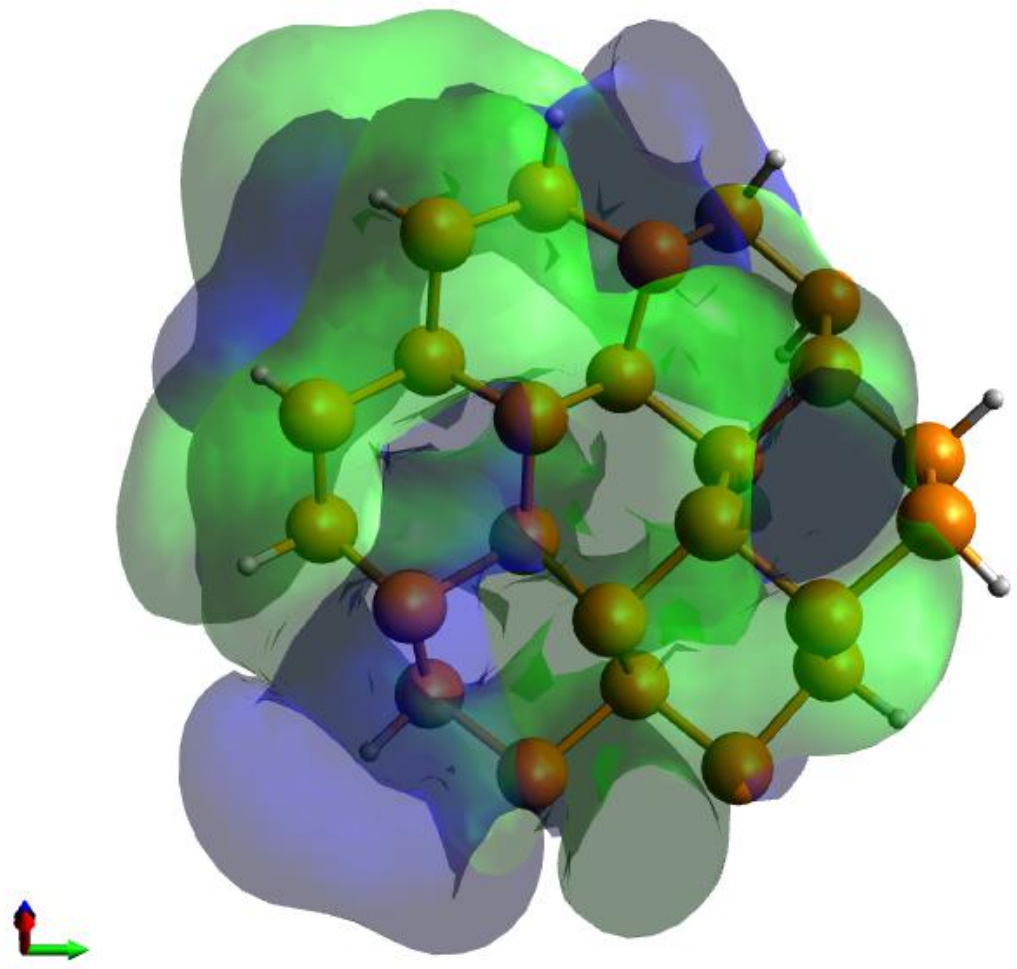

Orbital 187: LUMO+1

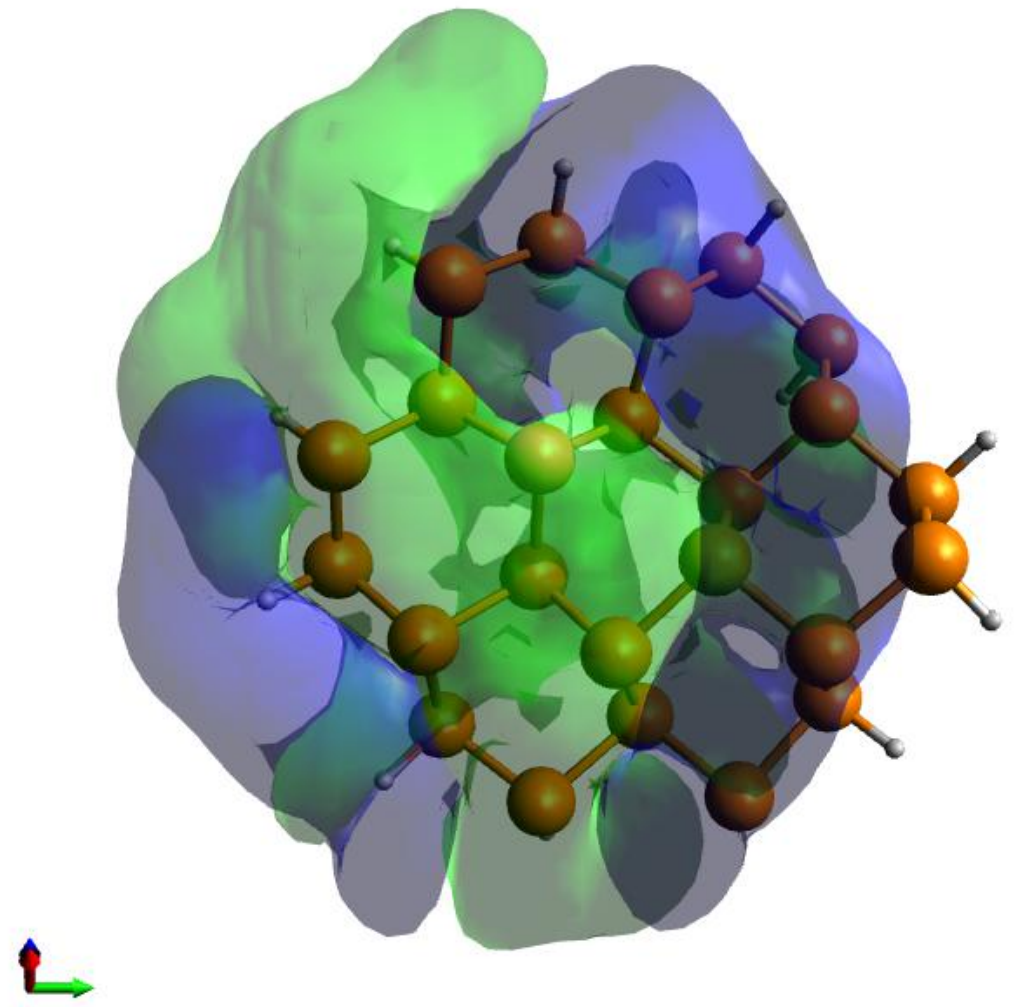

**- 4BenzP24H8**

States involved in the lowest transistion at 2.947 eV (420.7 nm).
Only excitations with weight larger than 0.01 are reported.

265a -> 266a : 0.917270
265a -> 267a : 0.034312
265a -> 268a : 0.012066

Orbital 265: HOMO

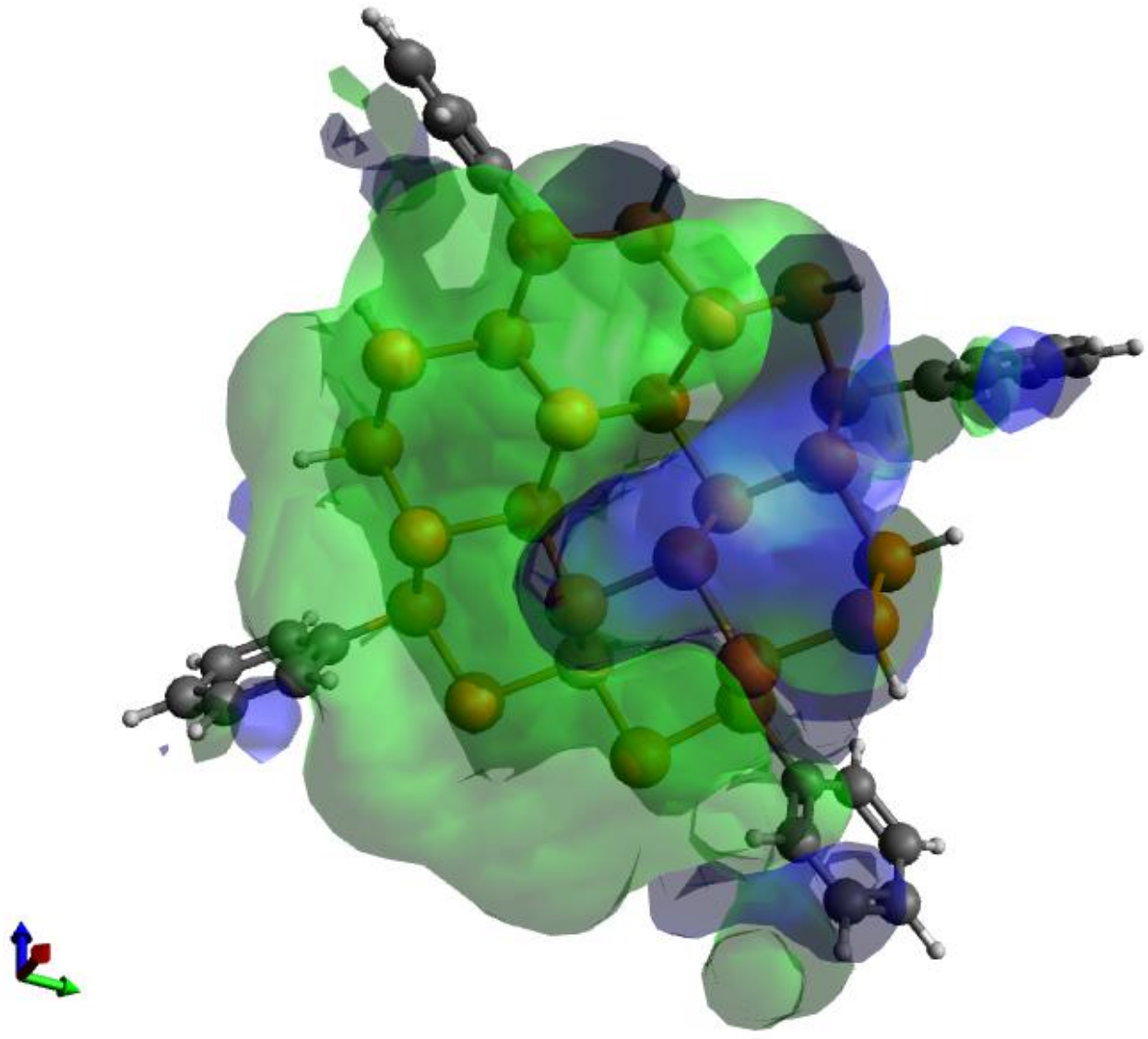

Orbital 266: LUMO

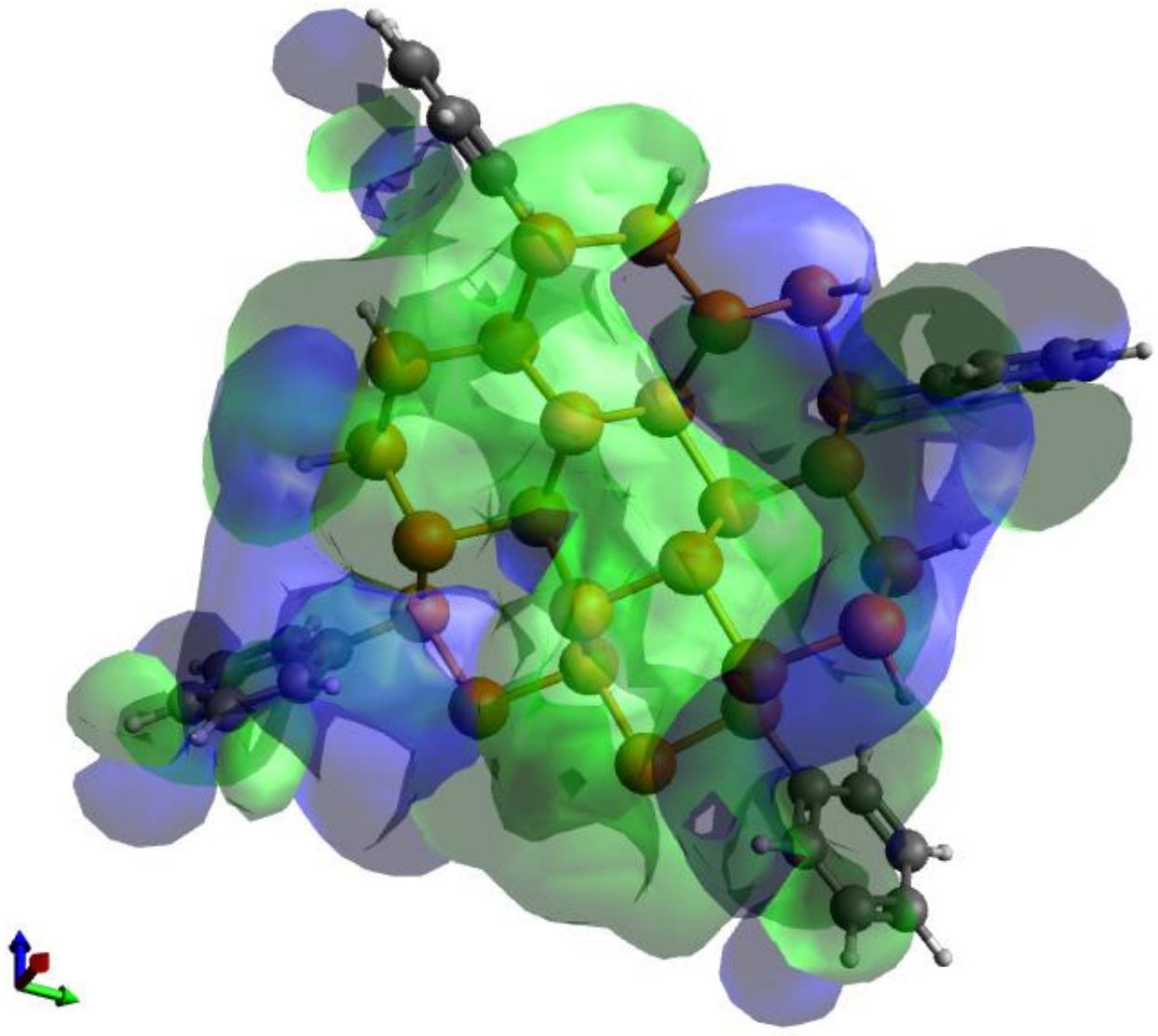

Orbital 267: LUMO+1

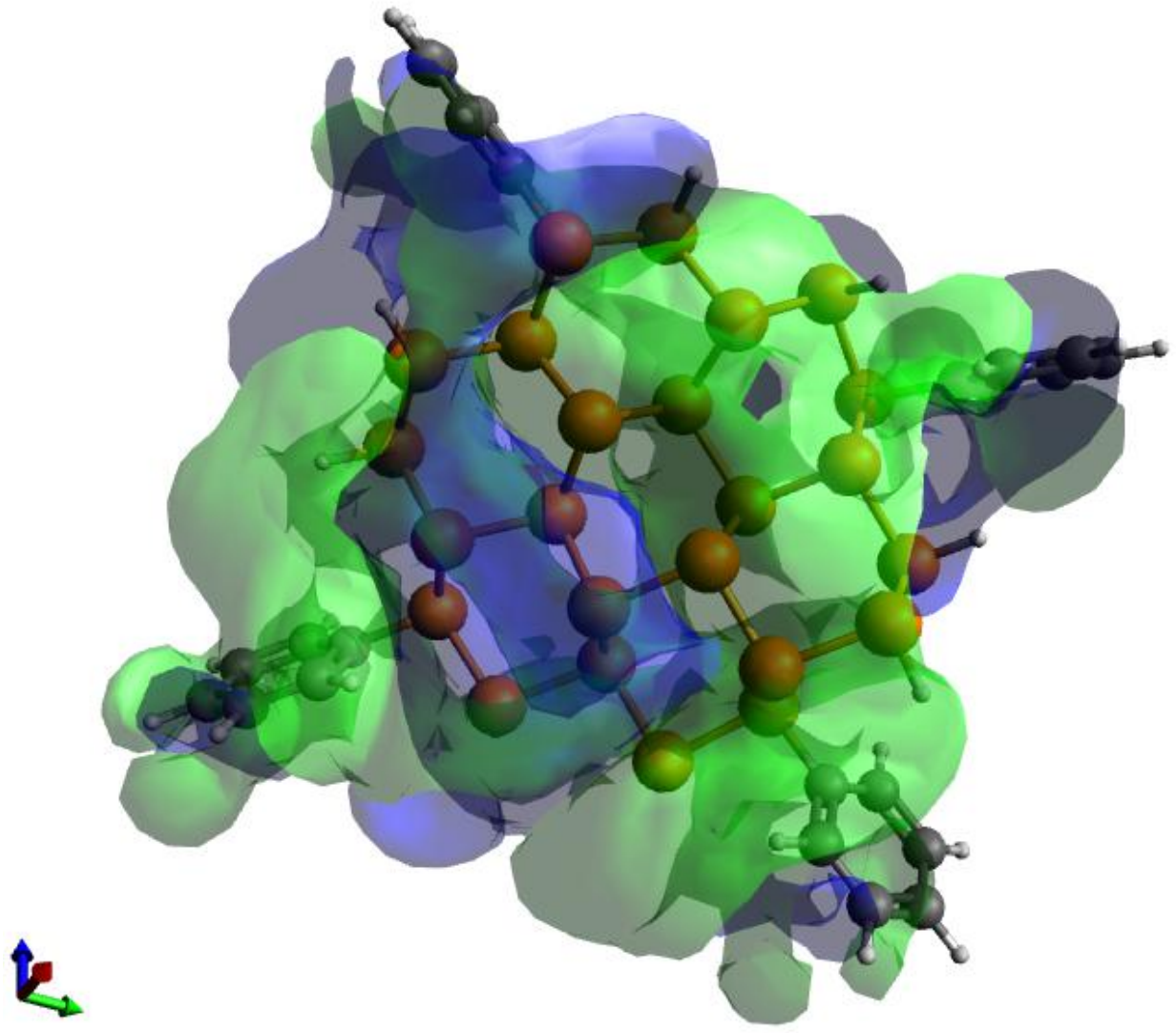

Orbital 268: LUMO+2

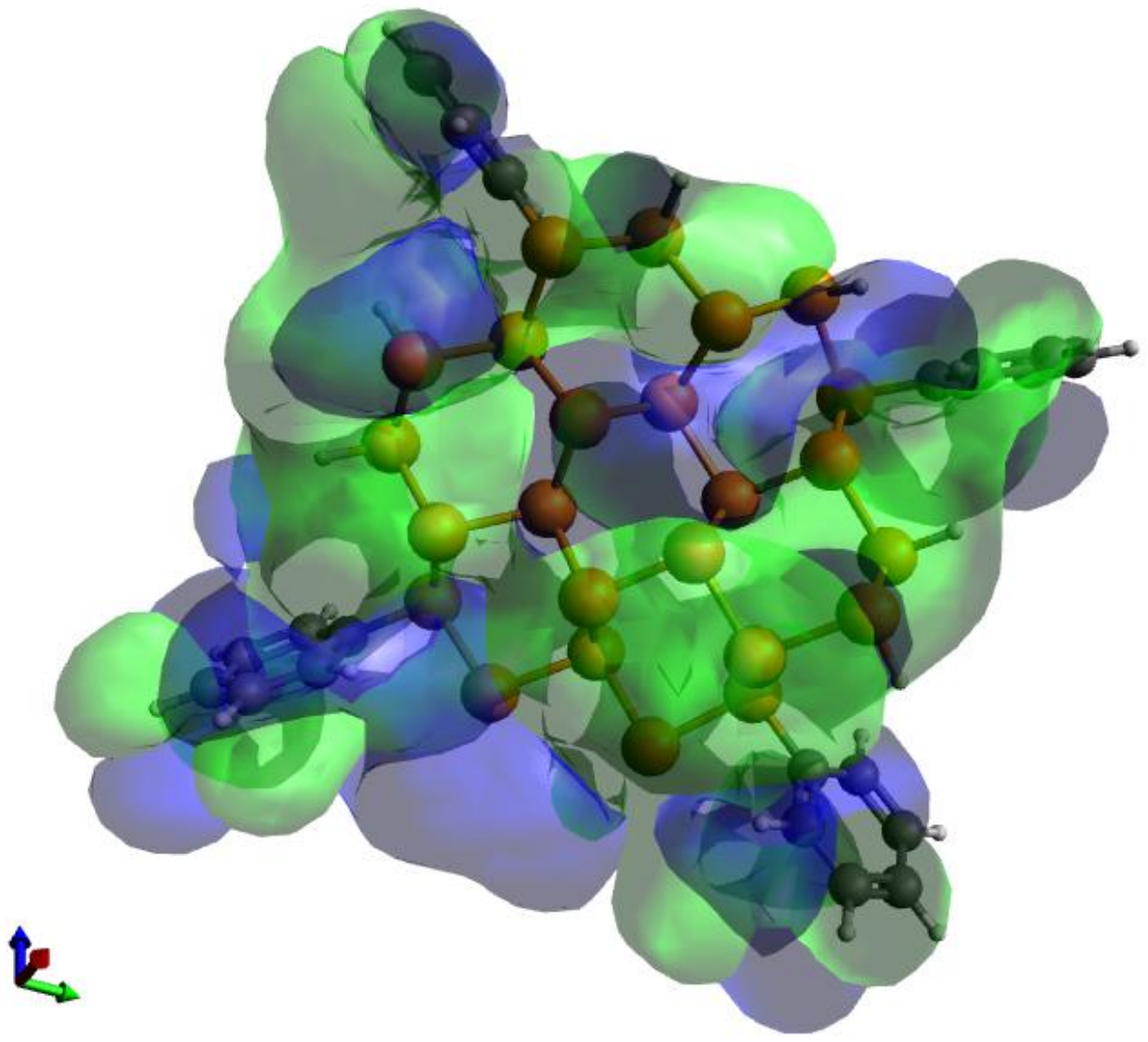

- **AnthP24H11**

States involved in the lowest transistion at 2.988 eV (415 nm).
Only excitations with weight larger than 0.01 are reported.
  230a -> 232a  :    0.140893
  230a -> 233a  :    0.824403

Orbital 230: HOMO

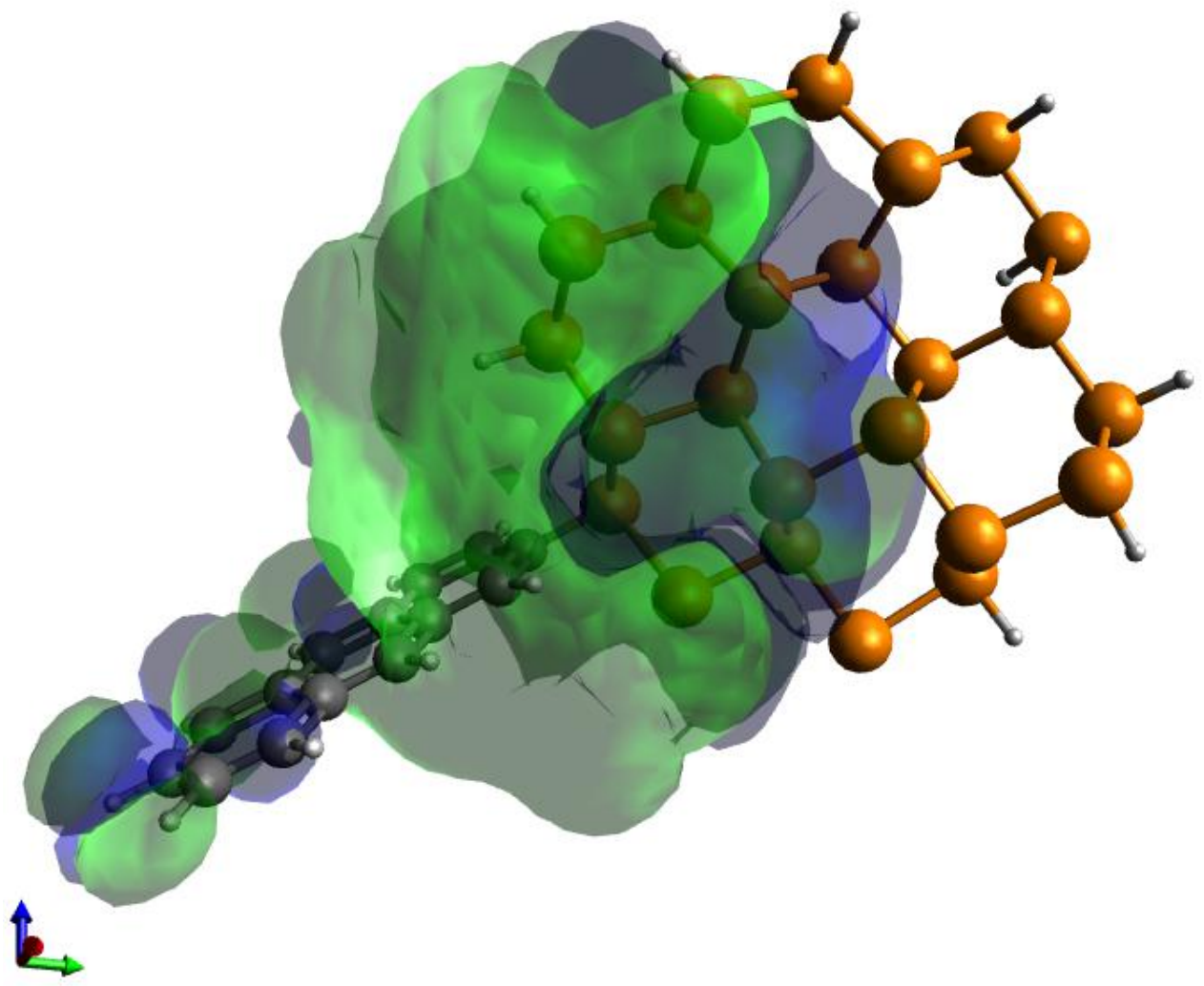

Orbital 233: LUMO+2

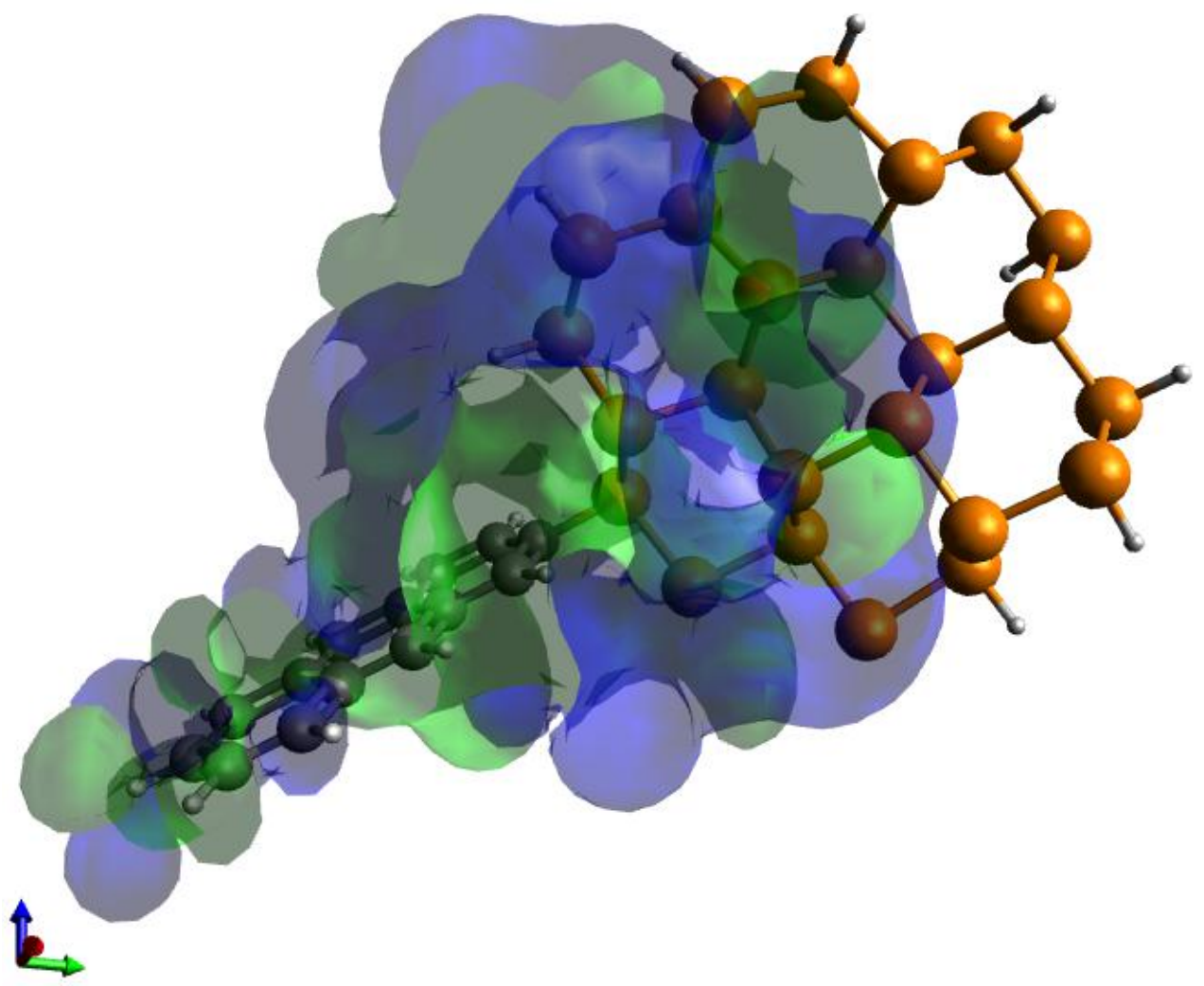

Orbital 232: LUMO+1

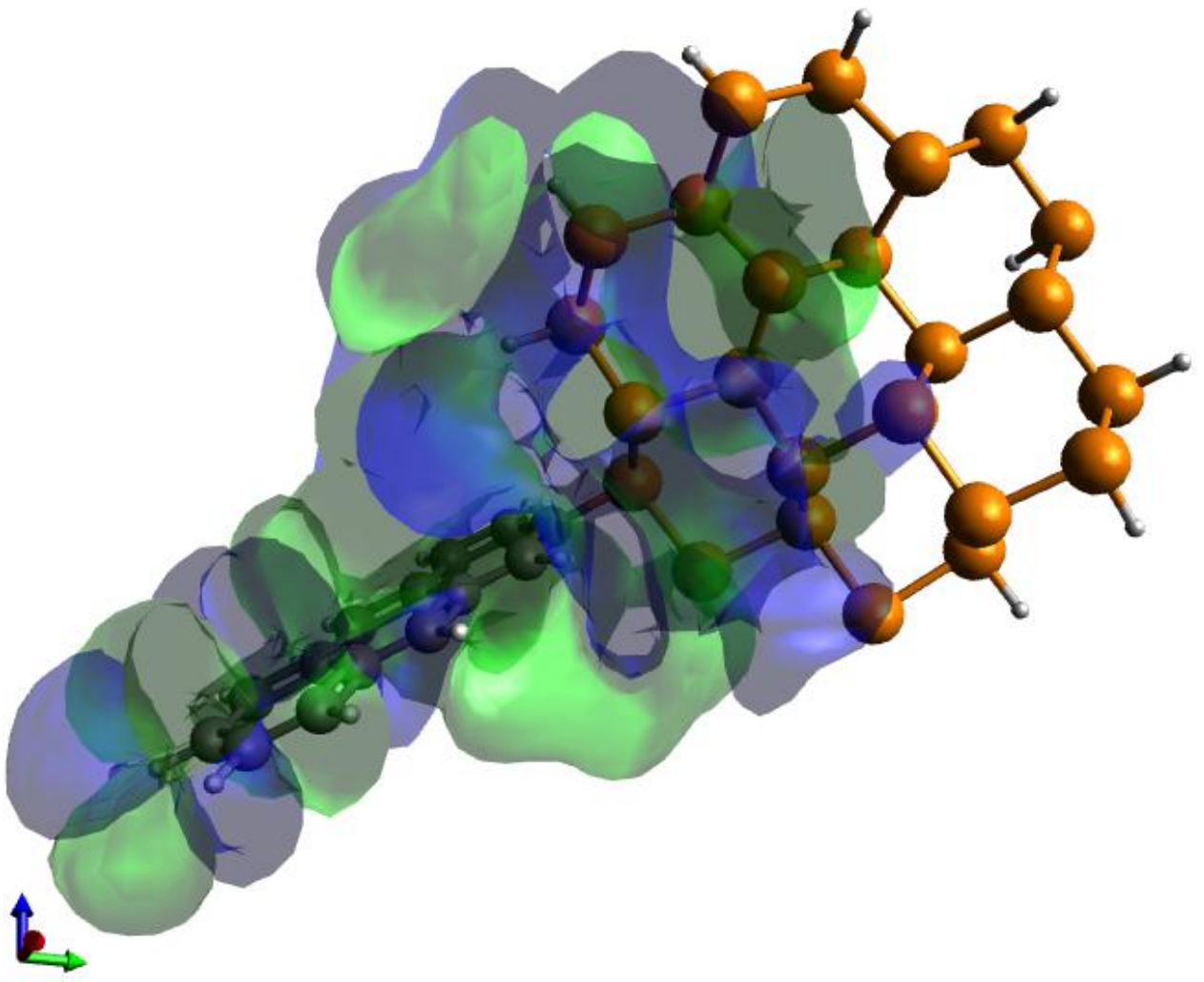